\documentclass[english,aip,letter,superscriptaddress,twocolumn,floatfix,jcp]{revtex4}
\usepackage[T1]{fontenc}
\usepackage[utf8]{inputenc}
\usepackage{xcolor}
\usepackage{graphicx}

\usepackage{subcaption}
\usepackage{amsmath,amssymb,amsfonts}
\usepackage{dsfont}
\usepackage[bookmarks=false]{hyperref}
\usepackage{calrsfs}

\newcommand{\niceT}{{\mathsf T}}

\DeclareMathAlphabet{\pazocal}{OMS}{zplm}{m}{n}
\newcommand{\calF}{\pazocal{F}}

\newcommand{\esc}{\!\cdot\!}
\newcommand{\llangle}{\left\langle}
\newcommand{\rrangle}{\right\rangle}

\begin{document}

\title{Self-averaging parameter   estimation    for
  coarse-grained particle models}

\author{Carlos Monago}
\author{J.A. de la Torre}
\author{Pep Español}
\email[]{pep@fisfun.uned.es}
\affiliation{Dept.  F\'{\i}sica  Fundamental, Universidad Nacional
  de   Educaci\'on   a   Distancia,  Spain}
\date{\today}

\begin{abstract}
  We introduce a parameter estimation method that utilizes microscopic
  data, specifically averages and correlations of selected microscopic
  observables,   to  determine   the   parameters   of  a   stochastic
  differential equation  governing coarse-grained degrees  of freedom.
  The  method  is  not  limited  to static  parameters  found  in  the
  reversible part of the coarse-grained dynamics, such as those in the
  free energy function or potential of mean force, but also extends to
  dynamic  parameters, including  friction  coefficients.  The  method
  couples the stochastic differential equation with free parameters to
  dynamic  equations  for the  parameters.   The  coupled system  self
  averages, according  to Anosov-Kifer's theorem,  in such a  way that
  the  final state  of the  parameters gives  coincidence between  the
  microscopic  and mesoscopic  averages and  correlations of  selected
  observables.  The  method is validated  in two examples:  a Brownian
  particle in  a harmonic potential,  and a set of  Brownian particles
  interacting hydrodynamically with the Rotne-Prager-Yamakawa mobility
  tensor.  This latter case illustrates how the method can be used not
  only to  determine coefficients  but also state  dependent transport
  properties  -- in  this case,  the  position dependent  form of  the
  mobility  tensor.  The  parameter  estimation for  these two  models
  yields excellent  results.  Subsequently  we use the  methodology to
  study a bimodal-mass Lennard-Jones fluid for which we infer both the
  potential  of  mean  force  between  the  heavy  particles  and  its
  hydrodynamic mobility tensor.
\end{abstract}

\maketitle
\section{Introduction}

Constructing  accurate  coarse-grained  (CG) models  from  microscopic
simulations is one  of the central challenges  of multiscale modeling.
In such models  a reduced set of CG variables  is selected to describe
the system at mesoscopic scales, and their evolution is represented by
an  effective  dynamics.   When  the dynamics  of  the  coarse-grained
variables arises from  the cumulative effect of  many fast microscopic
processes, such as collisions or molecular vibrations, a separation of
time scales  emerges. The  slow evolution of  the coarse  variables is
then  accompanied by  fast  fluctuations that  can  be represented  as
stochastic noise. This leads to  an effective Markovian description in
terms of a stochastic differential  equation (SDE).  This SDE contains
both reversible  and dissipative contributions.  The  reversible terms
are typically  related to  gradients of  a CG  free energy,  while the
dissipative terms  involve friction  coefficients or  mobility tensors
that encode the dynamic influence of unresolved degrees of freedom.

The   connection   between   microscopic   dynamics   and   stochastic
coarse-grained  descriptions   has  a  long  history   in  statistical
mechanics. The microscopic basis of stochastic coarse-grained dynamics
can  be  traced  back  to   Einstein's  analysis  of  Brownian  motion
\cite{Einstein1905}.  Studying  a  particle   suspended  in  a  fluid,
Einstein  derived the  diffusion  equation and  related the  diffusion
coefficient to the underlying molecular motion. Green \cite{Green1952}
subsequently  generalized   this  framework   to  arbitrary   sets  of
coarse-grained variables,  while Zwanzig  \cite{Zwanzig1961} rederived
the same structure using projection-operator techniques.

Within  this  framework the  drift  and  diffusion coefficients  admit
formal   microscopic    expressions\cite{Green1952,Zwanzig1961}.    In
general these  quantities can  be written as  conditional expectations
with  respect to  the  microscopic distribution  constrained to  fixed
values of  the CG  variables. Although  formally exact,  the practical
evaluation  of these  expressions  is extremely  challenging when  the
coefficients  depend  on the  state  of  many  CG variables,  since  it
requires  sampling high-dimensional  conditional distributions.   This
difficulty   is   a  manifestation   of   the   well-known  curse   of
dimensionality.

For this  reason, practical  CG models typically  introduce parametric
representations for the drift and  diffusion terms.  The parameters of
these  representations must  then be  inferred from  microscopic data.
The predictive power of a CG model therefore depends critically on the
ability to  determine these  parameters in  a systematic  and reliable
way.   Developing  such  inference  procedures  remains  an  important
problem in statistical mechanics.

Substantial progress  has been achieved in  determining the reversible
structure of  CG models\cite{noid2023}.  A number  of well-established
approaches allow one to infer effective free energies or potentials of
mean force  from microscopic simulations.  Prominent  examples include
inverse Monte  Carlo techniques \cite{lyubartsev1995},  force matching
methods \cite{izvekov2005,noid2008}, and relative entropy minimization
\cite{shell2008}.  These  methods construct  CG models  that reproduce
equilibrium   distributions  and   structural   correlations  of   the
underlying   system.    In   contrast,  the   determination   of   the
\emph{dynamical} parameters  of CG models remains  considerably more  difficult when  the friction  is state
dependent
\cite{papavasiliou2009a,hijón2010,dequidt2015,han2021,sokhan2021,abdulle2023,milster2025}.
Recent  work  has  explored data-driven  identification  of  effective
stochastic        equations        using       machine        learning
\cite{dietrich2023,ye2023, sachs2025}.

In this  work we introduce  a parameter estimation method  that infers
both  reversible and  dissipative parameters  of CG  stochastic models
directly  from  microscopic  averages  and  correlations  of  selected
observables.  The central idea is to reformulate the inference problem
as a dynamical one.  Instead  of determining the parameters through an
external optimization procedure, we  couple the CG stochastic dynamics
to evolution equations for  the parameters themselves.  The parameters
evolve in time so as to reduce the discrepancy between the microscopic
statistics of the observables and those generated by the CG model.

The  resulting coupled  system  possesses  a self-averaging  property.
Under suitable  ergodicity conditions, the long-time  evolution of the
parameter  dynamics drives  the  system toward  values  for which  the
averages and correlations produced by the CG model coincide with those
obtained  from  the microscopic  dynamics.   This  convergence can  be
understood within  the general framework of  self-averaging properties
of stochastic dynamical systems described by the Anosov--Kifer theorem
\cite{kifer2001}.  In this way, the  parameters of the CG model emerge
as the stationary state of an extended stochastic dynamics.

This self-averaging approach  has recently been explored by  us in the
context  of  coarse-grained  protein  models  \cite{Monago2025},  with
constant friction coefficients.  The  framework developed here extends
this  idea  to  general  CG  stochastic  particle  models  with  state
dependent transport properties, and it provides a unified strategy for
determining both thermodynamic parameters  (such as potentials of mean
force)  and  dynamical  parameters   (such  as  friction  or  mobility
tensors).

The  present   approach  is  related  in   spirit  to  stochastic
  approximation  and  online  parameter estimation  methods,  such  as
  Robbins–Monro schemes,  in which parameters are  updated using noisy
  estimates of target  quantities. However, a key  distinction is that
  here the parameter dynamics is embedded directly into the stochastic
  evolution  of the  coarse-grained  variables, leading  to a  coupled
  fast–slow system. In this  formulation, parameter estimation emerges
  as part of the physical dynamics  itself, rather than as an external
  optimization  procedure,  and  its convergence  can  be  interpreted
  through averaging principles.
  
  To  validate   the  method   we  consider  examples   of  increasing
  complexity.   We first  analyze a  Brownian particle  in a  harmonic
  potential, which provides  a simple setting in  which the parameters
  can be determined analytically.  We  then study a system of Brownian
  particles      interacting     hydrodynamically      through     the
  Rotne--Prager--Yamakawa  mobility  tensor,   demonstrating  how  the
  method can  recover a position-dependent mobility  matrix.  In these
  first two cases the validation is performed in a controlled way.  We
  generate  reference  data  using   a  mesoscopic  model  with  known
  parameters  and  then treat  these  data  as  if their  origin  were
  unknown.  Applying  the self-averaging inference procedure  to these
  data allows us to recover the  parameters and verify that the method
  re-obtain the parameters used to generate the dynamics.

Having  validated the  approach in  this way,  we then  apply it  to a
genuine coarse-graining problem.  In this  case the reference data are
obtained  from molecular  dynamics  simulations, and  the  goal is  to
determine the parameters of a  CG stochastic model that represents the
microscopic system at the mesoscopic  level.  The system considered is
a Lennard--Jones mixture composed of  light and heavy particles.  This
LJ fluid  model finds motivation  in numerous problems in  biology and
soft matter, where  disparities in mass and time  scales are essential
\cite{Mason1997,waigh2005}.   From  the  microscopic  trajectories  we
infer both the potential of mean force between the heavy particles and
the  corresponding  hydrodynamic  mobility  tensor.   The  problem  of
obtaining the mobility tensor for the case of two heavy particles from
MD     simulations    was     considered    by     Bocquet    \emph{et
  al.}~\cite{bocquet1997} by  computing position  dependent Green-Kubo
expressions  that for  two spheres  are not  subject to  the curse  of
dimensionality.

These results demonstrate that the proposed self-averaging framework
provides a systematic route for determining both thermodynamic and
dynamical parameters of coarse-grained stochastic particle models
directly from microscopic simulations.

\section{Dynamic coarse-graining}\label{Sec:coarse}

Coarse  graining   aims  at  representing  a   many-body  system  with
substantially less detail than  its full microscopic description.  The
microscopic state  is denoted  by $z\in\mathbb{R}^{6N}$  and comprises
the positions and momenta of  all atoms.  The microstate $z_t$ evolves
according  to  Hamilton's  equations   generated  by  the  Hamiltonian
$\hat H(z)$. At mesoscopic or macroscopic scales the system is instead
described  by a  reduced set  of  variables that  retain only  partial
information  about the  microstate.   These variables  are defined  as
phase  functions $\hat  A(z):\mathbb R^{6N}\rightarrow\mathbb  R^{M}$,
and are  commonly referred  to as CG  variables. The
microscopic trajectory $z_t$ induces  a trajectory $\hat A(z_t)$ whose
numerical  values are  denoted  by $a_t$.   The  central objective  of
coarse-graining  theory is  to derive  closed equations  governing the
evolution  of  $a_t$ without  explicit  reference  to the  microscopic
state.  This  was  achieved  by  Green  \cite{Green1952}  and  Zwanzig
\cite{Zwanzig1961} that  formulated a  Fokker-Planck equation  for the
probability density $P(a,t)$
\begin{align}
\partial_t P(a,t)
=&
-\nabla\!\cdot\!\big(D^{(1)}(a)\,P(a,t)\big)
  \nonumber\\
  &+\frac{1}{2}\nabla\nabla : \big(D^{(2)}(a)\,P(a,t)\big).
\label{FPE}
\end{align}
with corresponding Ito SDE
\begin{align}
da_t = D^{(1)}(a_t)\,dt + B(a_t)\,dW_t ,
\label{SDE}
\end{align}
where  $D^{(1)}(a)$  is the  drift  vector,  $W_t$ a  standard  Wiener
process, and $B(a)$  is the noise amplitude related to the diffusion
tensor $D^{(2)}(a) $ by the Fluctuation-Dissipation theorem
\begin{align}
D^{(2)}(a) = B(a)B^{\niceT}(a).
\label{D2BBT}
\end{align}
where $\niceT$ stands for matrix transpose.

As discussed in the Introduction, Green and Zwanzig derived formal
microscopic expressions for the drift vector and diffusion tensor in
terms of conditional equilibrium expectations. In practice, however,
the evaluation of these expressions becomes intractable when the
number of coarse-grained variables is large, due to the curse of
dimensionality. This limitation motivates the development of
alternative strategies that exploit microscopic information in a
computationally feasible way to infer the drift and diffusion
coefficients.
A  viable strategy  to  overcome  the curse  of  dimensionality is  to
propose    physically     motivated    or     sufficiently    flexible
\textit{parameterized      models},       $D^{(1)}(a,\theta)$      and
$D^{(2)}(a,\theta)$,
\begin{align}
  da_t
  &=D^{(1)}(a_t,\theta)dt+B(a_t,\theta)dW_t,
    \nonumber\\
   D^{(2)}(a,\theta)&=B(a,\theta)B(a,\theta)^T
  \label{SDE-gen}
\end{align}
The task is  then reduced to find the best  set of parameters $\theta$
that    represent     accurately    the    ``true''     CG    dynamics
(\ref{SDE}).

The  parameter  set  $\theta=(\lambda,\gamma)$   of  the  SDE  can  be
partitioned into  two distinct  subsets. The  \emph{static} parameters
$\lambda$   determine   the   equilibrium  distribution   sampled   by
Eq.~(\ref{SDE-gen}), defined as
\begin{align}
P^{\rm eq}_\lambda(a)
=
\frac{e^{-\beta \calF_\lambda(a)}}{Z_\lambda},
\label{Plambda}
\end{align}
where  $\calF_\lambda(a)$  is   the  associated  free-energy  function
defined through  this equation,  and $Z_\lambda$ is  the normalization
constant. This  model distribution  is constructed to  approximate the
``true'' equilibrium distribution,
\begin{align}
P_*^{\rm eq}(a)
=
\int dz\, \rho_*^{\rm eq}(z)\,
\delta\!\left(\hat A(z)-a\right),
\label{Pmic}
\end{align}
The underlying equilibrium measure  $\rho_*^{\rm eq}(z)$ is assumed to
be canonical,
\begin{align}
\rho_*^{\rm eq}(z)
=
\frac{e^{-\beta \hat H(z)}}{Z(\beta)}.
\label{Gibbs}
\end{align}
where $Z(\beta)$ is the partition function.
It is convenient to introduce the
free-energy function $\calF_*(a)$ through
\begin{align}
P_*^{\rm eq}(a)
=
\frac{e^{-\beta {\calF}_*(a)}}{Z_*},
\label{pal}
\end{align}
which defines ${\calF}_*(a)$  up to an additive  constant. Good static
parameters should lead to $\calF_\lambda(a)\simeq\calF_*(a)$. In this work
we take starred symbols as ``ground truth''.

In  contrast,  the  \textit{dynamic}  parameters  $\gamma$,  typically
friction  or diffusion  coefficients, do  not affect  the form  of the
equilibrium distribution function by construction, as described above.
We discuss the estimation of both types of parameters separately.

\section{Estimation of static parameters $\lambda$}
In  order  to  estimate  the   parameters  $\lambda$  that  enter  the
equilibrium  probability  (\ref{Plambda})  a  common  strategy  is  to
consider a set  of \textit{CG observables}, i.e., functions  of the CG
variables,   and   require   that   their   averages   computed   with
$P_\lambda^{\rm  eq}(a)$   reproduce  the   corresponding  microscopic
averages  obtained  from atomistic  MD  simulations.   In CG  particle
methods where bunchs of atoms are grouped into CG particles, different
methods  implement  this idea  by  choosing  different observables  to
match.  In  the Relative Entropy  method \cite{shell2008}, the
parameters   are  determined   by  minimizing   the  Kullback--Leibler
divergence between the microscopic  distribution projected onto the CG
variables and  the model  distribution $P_\lambda^{\rm  eq}(a)$.  This
condition  is equivalent  to  matching the  ensemble  averages of  the
observables $\partial  U_\lambda/\partial \lambda$,  where $U_\lambda$
is the parameterized  potential of mean force, in  the microscopic and
CG  ensembles. In  the  force matching  or multiscale  coarse-graining
method \cite{noid2008}, the observables whose averages are matched are
the forces acting on the CG  variables: the parameters are obtained by
minimizing  the mean-square  difference between  microscopic projected
forces and the CG forces. Structure-based approaches such as iterative
Boltzmann   inversion   \cite{Reith2001}  instead   match   structural
observables, typically  pair correlation functions, by  requiring that
the radial distribution  functions produced by the  CG model reproduce
those measured in the atomistic system.

Let   us   consider  a   set   of   $L$ CG   observables  $O(a)   \in
\mathbb{R}^L$.   The    equilibrium   average   of    the   observable
$O(\hat{A}(z))$ computed microscopically is denoted as
\begin{align}
  \label{Omic}
\llangle O\rrangle_*=\int dz\,\rho_*^{\rm eq}(z)\, O(\hat{A}(z)).
\end{align}
Using (\ref{Pmic}) this can be written as
\begin{align}
\llangle O\rrangle_*=\int da\, P_*^{\rm eq}(a)\,O(a),
  \label{OmiPmic}
\end{align}
This is taken as the ``ground  truth''. On the other hand, the average
of the  CG observables with  the parameterized model  (\ref{Plambda}) are
denoted with
\begin{align}
  \llangle O\rrangle_\lambda\equiv\int da\, P_\lambda^{\rm eq}(a)\,O(a).
  \label{oOlambda}
\end{align}
The goal is to find the set of $P$ parameters $\lambda$ that guarantee the
alignment of the  equilibrium averages of $O(a)$  computed both, using
the  CG model and  using the  microscopic  dynamics. In  other
words, our objective is to attain the following relationship,
\begin{align}
  \label{oFFmic}
\llangle O\rrangle_\lambda=  \llangle O\rrangle_* .
\end{align}

Observe that (\ref{oFFmic})  is a set of $L$  non-linear equations for
the  parameters  $\lambda$. In general, we expect that  if $L=P$ we
may have  a unique  solution, although this  is not  always guaranteed
(see     the      nice     discussion     in     Chapter      9     of
\cite{Press1992}).  Nevertheless, for  the time  being we  will assume
that $L=P$,  that is, we choose  as many CG observables  as parameters
$\lambda$ in our parameterized model.

To solve  (\ref{oFFmic}) for $\lambda$ we may  use the iterative
  Newton-Raphson method.  The iterative  nature of the method suggests
  to introduce a continuum evolution in parameter space of the form

\begin{align}
\dot{\lambda}_t &=-\frac{1}{T_{\rm param}}\left[ \llangle O\rrangle_*  -\llangle O\rrangle_{\lambda_t} \right],
                  \label{ldyn}
\end{align}
where $\dot{\lambda}_t=\frac{d\lambda_t}{dt}$ and $T_{\rm param}$ is a
convenient time scale. The fixed point of  this dynamical system
  gives (\ref{oFFmic}).

Our      proposal      is       to      perform      the      averages
$\llangle\cdots\rrangle_\lambda$ required in Eq.  (\ref{ldyn})
{\em at the same time} as  the evolution of $\lambda$ takes place \cite{Monago2025}.  To
this end,  we {\em enlarge}  the state  space of the  coarse variables
with the set of parameters  and propose the following coupled dynamics
in the enlarged space
\begin{align}
  da_t&=D^{(1)}(a_t,\theta_t)dt+B(a_t,\theta_t)dW_t,
\nonumber\\
\dot{\lambda}_t &=-\frac{1}{T_{\rm param}}\left[ \llangle O\rrangle_*  -O(a_t)\right].
\label{aldyn}
\end{align}
The   first   equation   is    just   the   coarse-grained   evolution
(\ref{SDE-gen})  where, for  the  time being,  the dynamic  parameters
$\gamma$ within the set $\theta$ are  assumed to be known.  The second
equation is suggested  by the form of Eq.   (\ref{ldyn}).  However, as
opposed to  Eq.  (\ref{ldyn}),  the key  observation is  that \textit{
  there  are  no  averages $\llangle  \cdots\rrangle_\lambda$  in  Eq.
  (\ref{aldyn}).}  The idea is to run the mesoscopic simulation with a
set of slowly  varying parameters $\lambda$ that evolve in  such a way
as to reduce as much as  possible the differences between the targeted
equilibrium  average from  the  actual average.   Therefore, the  time
scale $T_{\rm param}$ that governs the scale of evolution of $\lambda$
is  a  time  scale  which  should  be  ``large''.   We  may  think  of
$T_{\rm param}$ as  the time span over which the  required averages in
Eq.  (\ref{ldyn}) are reliably computed through time averages.

In order to be more precise and see that Eq. (\ref{aldyn}) will do the
job, we  rescale time  with $\epsilon\equiv T_{\rm param}^{-1}$
and rewrite
Eq.  (\ref{aldyn})  in  a  standard  way  in  the  analysis  of  stiff
differential equations
\begin{align}
\dot{\lambda}_t &=-\Big(\llangle O\rrangle_* -O(a_t)\Big),
\nonumber\\
d  a_t &=\frac{1}{\epsilon} D^{(1)}(a_t,\theta_t)\,dt+\frac{1}{\sqrt{\epsilon}}B(a_t,\theta_t)\,dW_t.
\label{aldyn2}
\end{align}
Equivalently, defining
\begin{align}
  g(a,\lambda)\equiv-\Big(\llangle O\rrangle_* -O(a)\Big)=O(a)-\llangle O\rrangle_* ,
  \label{g}
\end{align}
the slow equation reads $\dot\lambda_t=g(a_t,\lambda_t)$.

In the  limit $\epsilon\to0$,  the variables  $ a $  are very  fast as
compared  with  the variables  $\lambda$.   We  will assume  that  the
dynamics of the fast variables $a$ is ergodic, implying that for every
fixed  $\lambda$  the  fast  equation  in  (\ref{aldyn2})  has  the
equilibrium  measure $P^{\rm  eq}_\lambda(a)$.   Now,  according to  a
general  theorem  of  averaging  due to  Anosov  \cite{Givon2004}  and
generalized to  the stochastic  realm by Kifer  \cite{kifer2001}, that
assumes the dynamics of $ a  $ is ergodic, in the limit $\epsilon\to0$
the  slow  variables  $\lambda$  evolve  according  to  the  following
\textit{closed} equation
\begin{align}
  \dot{\lambda}_t &=G(\lambda_t)
\label{lcdyn}
\end{align}
where the function $G(\lambda)$ is defined as
\begin{align}
  G(\lambda) & \equiv  \int d a \, P_\lambda^{\rm eq}( a)\, g( a,\lambda )
=\llangle O\rrangle_\lambda-\llangle O\rrangle_* .
\end{align}
In order  to close  the argument,  note that  if the  dynamic equation
(\ref{aldyn2})   has  a   stationary  state   $\lambda^*$  for   which
$\dot{\lambda}_t=0$,   then,   the   values   $\lambda^*$   satisfies
$G(\lambda^*)=0 $. By using (\ref{g}), the condition $G(\lambda^*)=0 $
is equivalent to
\begin{align}
  \label{l*}
  \llangle O\rrangle_{\lambda^*} &=\llangle O\rrangle_* .
\end{align}
Therefore, the stationary  value $\lambda^*$ at which  the dynamics of
$\lambda_t$ converges is  precisely the one that ensures  that the CG
model has the real averages for the CG observables.

In summary,  we solve the problem  of finding the root  $\lambda^*$ of
the non-linear equation (\ref{oFFmic}) by running a sufficiently long
but  inexpensive coarse-grained  simulation  in  which the  parameters
$\lambda$ of the  free energy depend on  time and evolve
according   to   Eqs.   (\ref{aldyn}).    In   the   limit  of   large
$T_{\rm param}$ and  for large times, the  parameters $\lambda$ should
converge towards the value $\lambda^*$ that gives the correct matching
(\ref{oFFmic}).
The fixed point $\lambda^*$ selected by this dynamics need not be unique
in general.  As in any root-finding procedure, if several stable fixed
points exist, the limiting value may depend on the basin of attraction
of the initial condition $\lambda_0$.  For parametrized equilibrium
distributions of exponential-family form, with observables conjugate to
the parameters, the matching condition~(\ref{oFFmic}) is equivalent to
minimizing the Kullback--Leibler divergence between the reference and
model distributions~\cite{shell2008}.  In that case, the corresponding
objective is convex in the natural parameters, so metastable local
minima are not expected when the parametrization is non-redundant and
the target averages can be represented by the model.  Outside this
setting, multiple fixed points may in principle exist.  In the examples
considered below, the physically motivated parametrizations employed
lead to a stable fixed point, which is validated a posteriori by the
observed convergence of the parameters and by the reproduction of
observables during the parameter evolution.%

The validity of the averaging argument relies on standard assumptions of fast–slow stochastic systems: (i) ergodicity of the coarse-grained dynamics for fixed parameters, (ii) sufficient separation between the time scales of the variables and the parameters, and (iii) smooth dependence of the dynamics on the parameters. While a rigorous verification of these conditions is beyond the scope of this work, they are expected to hold in the regimes considered here and are supported a posteriori by the observed convergence of the parameter dynamics.

\section{Estimation of dynamic parameters $\gamma$}
Dynamic    parameters     can    be    estimated     from    comparing
\textit{equilibrium}   time   auto-correlation   functions   (at   one
particular   time    $\Delta   t$)   for   some    other   observables
$Q_\alpha(a), \alpha=1,\cdots,P$.  The  $P$ auto-correlation functions
are defined microscopically as
\begin{align}
  \llangle Q_\alpha(a_0)Q_\alpha(a_{\Delta t})\rrangle_*
  &=\int dz \,\rho^{\rm eq}_*(z)\,Q_\alpha(\hat{A}(z))\,Q_\alpha(\hat{A}(z_{\Delta t}))
    \label{Cmic}
\end{align}
where $z_{\Delta t}$  is the microstate at time ${\Delta  t}$ when the
initial microstate  is $z$.  The correlation  function (\ref{Cmic}) is
measured  in an  MD simulation  through time-origin  averages, and  is
taken as the ground truth.

By analogy with the static case, we require the parameters $\theta$ satisfy the non-linear equation
\begin{align}
\llangle Q_\alpha(a_0)\,Q_\alpha(a_{\Delta t})\rrangle_{\theta}
 \simeq
\llangle Q_\alpha(a_0)\,Q_\alpha(a_{\Delta t})\rrangle_*
\label{C-mes-mic}\end{align}
where $\llangle\cdot\rrangle_\theta$ denotes  a stationary time-origin
average  under the  CG  dynamics \eqref{SDE-gen}  at fixed  parameters
$\theta$. Observe that  this equation is to  be satisfied \textit{only
  at one specific value of the time lag $\Delta t$}.

To  find the  value  of  the parameters  $\theta$  that fulfill  this
condition,  we  propose  to  couple the  SDE  (\ref{SDE-gen})  to  the
following dynamics for the parameters $\theta=(\lambda,\gamma)$,
\begin{align}
  da_t&=D^{(1)}(a_t,\theta_t)\,dt+B(a_t,\theta_t)\,dW_t
             \nonumber\\
   \dot{\lambda}(t)
  &= -\frac{1}{T_{\rm param}}\left[\llangle O\rrangle_*-O(a_t)\right]
    \nonumber\\
  \dot{\gamma}(t)
  &= -\frac{s_Q}{T_{\rm param}}\left[C_{*}(\Delta t)-\widehat C(t;\Delta t)\right],
\label{dalg}
\end{align}
where $C_{*}(\Delta t)\equiv \llangle Q(a_0)Q(a_{\Delta t})\rrangle_*$
is  the   target  vector  of  microscopic   auto-correlations  at  the
particular time  lag $\Delta  t$, and $\widehat  C(t;\Delta t)$  is an
instantaneous estimator computed along the CG trajectory.  To make the
update causal, we use the lagged estimator
\begin{equation}
  \widehat C(t;\Delta t)\equiv Q(a_t)\,Q(a_{t-\Delta t}),
  \label{Cinst}
\end{equation}
implemented in  practice by  storing in  a buffer  a short  history of
$Q(a)$ values  over the lag  $\Delta t$.  Here $s_Q\in\{+1,-1\}$  is a
sign that depends  on the chosen correlation (and, in  general, on the
parametrization of $\gamma$), and is  fixed by the stability criterion
derived  in Appendix \ref{App:sign}.

We  expect  that   the  dynamics  of  the   system  (\ref{dalg}),  for
sufficiently  large $T_{\rm  param}$,  will lead  to a  self-averaging
property  in  which   $\theta_t=(\lambda_t,\gamma_t)$  obey  a  closed
equation of the form
\begin{align}
     \dot{\lambda}(t)
  &= -\frac{1}{T_{\rm param}}\left[\llangle O\rrangle_* -\llangle O\rrangle_{\lambda_t}\right]
    \nonumber\\
  \dot{\gamma}(t)
  &= -\frac{s_Q}{T_{\rm param}}\left[C_{*}(\Delta t)- C(\Delta t;\theta_t)\right].
\end{align}
where $C(\Delta t;\theta_t)=\llangle Q(a_t)Q(a_{t-\Delta t})\rrangle_\theta$.
In the limit $t\to\infty$ we expect  this set of equations to converge
to values $\theta^*=(\lambda^*,\gamma^*)$ that ensure
\begin{align}
\llangle O\rrangle_{\lambda^*}
  &=  \llangle O\rrangle_* 
    \nonumber\\
  C(\Delta t;\theta^*)
  &=C_{*}(\Delta t).
    \label{converged}
\end{align}

We require  that the microscopic and  mesoscopic correlation functions
coincide  at a  single time  lag, \(  \Delta t  \).  This  provides an
\textit{a posteriori}  validation of the  CG model.  On one  hand, the
estimated parameters \(  \lambda^* \) and \( \theta^*  \) will depend,
in general,  on the choice of  \( \Delta t  \).  If, and only  if, the
parametric  form  of  the  SDE accurately  represents  the  underlying
all-atom  dynamics,  these   parameters  should  remain  approximately
independent of \( \Delta t \). Therefore,  the range of \( \Delta t \)
over which the estimates are invariant provides a consistency check of
the model.   On the other  hand, an additional validation  consists in
verifying that the time correlation  functions \( C(t;\theta^*) \) and
\( C_{*}(t) \) agree over the full range of times, and not only at the
particular value $t=\Delta t$ fixed by construction.

Finally,  the  choice  of  observables used  to  drive  the  parameter
dynamics is  not unique and should  be guided by the  structure of the
parametrized model. In practice, observables are selected so that they
are  sensitive  to  the  parameters of  interest  and  reflect  either
equilibrium  or  dynamical properties  of  the  system.  A  systematic
analysis of  identifiability and optimal observable  selection is left
for future work.

\begin{figure*}[t]
  \centering

  \begin{subfigure}[t]{0.49\linewidth}
    \includegraphics[width=\linewidth]{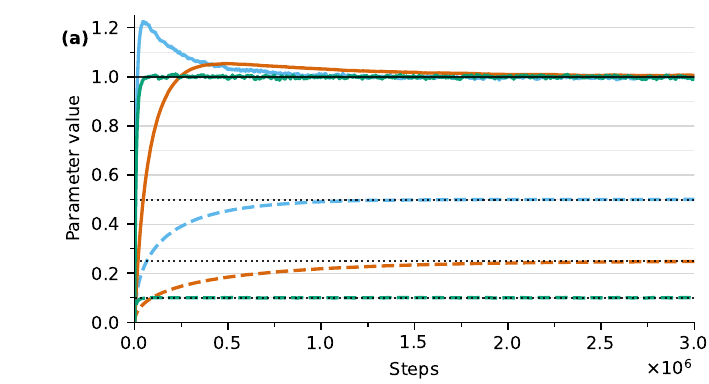}
  \end{subfigure}\hfill
  \begin{subfigure}[t]{0.49\linewidth}
    \includegraphics[width=\linewidth]{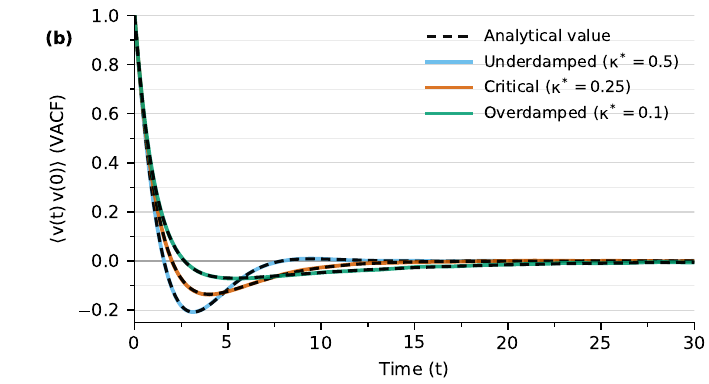}
  \end{subfigure}
  \caption{Langevin  particle in  a harmonic  trap in  three dynamical
    regimes.    \textbf{(a)}   Parameter   relaxation  of   a   single
    realization under the self-averaging  dynamics for the underdamped
    ($\kappa^\ast=0.5$), critical ($\kappa^\ast=0.25$), and overdamped
    ($\kappa^\ast=0.1$)  regimes.  Solid   lines  represent  the  time
    evolution of  the friction  coefficient $\gamma(t)$,  while dashed
    lines correspond to the spring constant $\kappa(t)$.  Horizontal
    reference lines  indicate the  target values  ($\gamma^\ast=1$ and
    $\kappa^\ast$).   \textbf{(b)}  Velocity autocorrelation  function
    $\langle   v(t)v(0)\rangle$   for   the   corresponding   regimes:
    simulation results  (solid colored)  compared with  the analytical
    prediction (black dashed).  The agreement is excellent.}
  \label{Fig:particleEvolution}
\end{figure*}

\section{Example 1: Brownian trap}
\label{B-1D}

In order to validate the  self-averaging method, we first consider the
following  simple SDE  for  the position  $q$ and  momentum  $p$ of  a
one-dimensional  Brownian  particle  in   a  harmonic  potential.  The
Langevin equation governing its dynamics is
\begin{align}
  \label{Langevin}
  dq& = \frac{p}{m}\, dt
  \nonumber\\
  dp &=-\kappa q\, dt -\frac{\gamma}{m} p\, dt +(2k_BT\gamma)^{1/2}\, dW_t,
\end{align}
where $m$ is the particle mass, $\kappa$ the spring constant, $\gamma$
the friction  coefficient, and  $T$ the  temperature. The  term $dW_t$
denotes an increment of  a Wiener process.  Equations (\ref{Langevin})
are an example of the  parameterized CG SDE (\ref{SDE-gen}), where the
CG    variables   are    $a=\{q,p\}$    and    the   parameters    are
$\theta=\{\kappa,\gamma\}$. While one could  also treat $m$ and $k_BT$
as parameters to be targeted, we assume them to be known.  We estimate
$\theta$  by comparing  the predictions  of (\ref{Langevin})  with the
``ground  truth''  provided  by  exact  analytical  results  given  in
Refs.~\cite{beard2000,tanygin2024}.

The equilibrium distribution, which is a solution of the FPE
corresponding to (\ref{Langevin}), is
\begin{align}
  P^{\rm eq}(q,p)=\frac{1}{Z}\exp\left\{-\frac{1}{k_BT}\left(\frac{p^2}{2m}+\frac{\kappa}{2}q^2\right)\right\}.
  \label{Equilibrium-Langevin}
\end{align}
The parameter $\kappa$ enters the equilibrium probability distribution
and is therefore a static parameter. By contrast, the friction $\gamma$
does not appear in (\ref{Equilibrium-Langevin}) and is a dynamic parameter. The covariances associated with (\ref{Equilibrium-Langevin}) are
\begin{align}
  \label{miccov}
  \llangle q^2\rrangle
  &= \frac{k_BT}{\kappa}
  \nonumber\\
  \llangle p^2\rrangle
  &= mk_BT.
\end{align}

As CG observables, we choose
\begin{align}
  O(a)&\to q^2
  \nonumber\\
  Q(a)&\to p(t)p(t-\Delta t ) 
\end{align}
In this  way, we  couple the SDE  (\ref{Langevin}) with  the following
equations for  the evolution of  the parameters $\kappa$  and $\gamma$,
with  the sign  convention discussed  in the  Appendix \ref{App:sign},
\begin{align}
  \label{dkdg}
  \frac{d\kappa(t)}{dt} &=-\frac{1}{T_{\rm param}}\left[\frac{k_BT}{\kappa}-q^2(t)\right]
  \nonumber\\
  \frac{d\gamma(t)}{dt} &=-\frac{1}{T_{\rm param}}\left[C_p(\Delta t )-p(t-\Delta t )p(t)\right].
\end{align}
where the momentum time autocorrelation function is
\begin{align}
  C_{p}(\Delta t )
  &\equiv \llangle p(0)p(\Delta t )\rrangle,
\label{Cp-tau}\end{align}
given analytically in Refs.~\cite{beard2000,tanygin2024}.
The correlation function of the parameterized model
\begin{align}
  C_\gamma(\Delta t )
  &\equiv \llangle p(0)p(\Delta t )\rrangle_\gamma,
\end{align}
will be compared with (\ref{Cp-tau}). 

Observe that  the procedure may be  implemented in two stages.  In the
first  stage, one  may couple  the evolution  equation for  the static
parameter $\kappa$ while  keeping $\gamma$ fixed at  an arbitrary (but
numerically   reasonable)  value.   This  is   possible  because   the
equilibrium   distribution  (\ref{Equilibrium-Langevin}),   where  the
static parameter appears, does not depend on the value of the friction
coefficient. Once the fitted value $\kappa^{*}$ has been obtained, the
second  stage   consists  of  coupling  the   evolution  equation  for
$\gamma$.  We have  verified,  however, that  coupling both  equations
(\ref{dkdg}) simultaneously to the SDE  yields the same result as this
two–stage procedure.

The simulation  details are  given in  Appendix \ref{App:sim-details}.
In  Fig.~\ref{Fig:particleEvolution}   both  parameters   are  evolved
simultaneously, starting from the  initial values $\kappa=10^{-3}$ and
$\gamma=10^{-3}$.   The figure  shows  the temporal  evolution of  the
parameters   according   to   Eq.~(\ref{dkdg}).   As   expected,   the
self-averaging   procedure    converges   to   the    correct   values
$\kappa=\kappa^\ast$   and  $\gamma=\gamma^\ast$   across  the   three
dynamical regimes  (underdamped, critical, and overdamped),  with high
accuracy. The  converged parameters reproduce the  analytical velocity
autocorrelation function (VACF)  over the entire time  window shown in
Fig.~\ref{Fig:particleEvolution}(b).

We take the  fitting lag as $\Delta t =n_{\Delta  t}\,dt$ with integer
$n_{\Delta t}$, and evaluate $p(t-{\Delta t})p(t)$ using a ring buffer
storing the past $n_{\Delta t}$ values of $p(t)$. We also checked that
different choices  of the  lag time  ${\Delta t}$  in Eq.~(\ref{dkdg})
lead  to the  same  fitted value  of $\gamma$.  In  practice, we  find
lag-independence  for ${\Delta  t}$ ranging  from $0.05$  up to  $10$,
$15$,  and $25$  (in time  units) for  the underdamped,  critical, and
overdamped regimes, respectively. In other words, essentially the full
range of  time lags $\Delta t$  in the VACF  can be used to  infer the
irreversible   parameter,   consistent   with   our   lag-independence
hypothesis when the fitted CG model matches the ground-truth physics. Even in  this favorable  case, it  is advisable to  choose a  lag time
inside a reasonable window (e.g.,  ${\Delta t}$ of order $1$--$2$ time
units),  both to  avoid  storing  a large  history  of velocities  and
because long  lags provide poorer  statistics.

\section{Example 2: Brownian trap with HI}
\label{N-B-HI}
A      second,       more      challenging       example      involves
\textit{position-dependent} frictions, as  those appearing in Brownian
Dynamics  with hydrodynamic  interactions (HI) or  in Dissipative  Particle
Dynamics. We consider  the first model, where  $N$ identical spherical
particles are suspended  in a fluid.  The system  is coarse-grained at
the level  of the particle positions  $\{{\bf R}_\mu\}_{\mu=1}^N$. The
SDE governing the particle positions is
\begin{align} \label{SDE-HI}
  d{\bf R}_\mu
  &=
  -\sum_\nu \mathbf{D}_{\mu\nu}\frac{1}{k_B T}\frac{\partial V^{MF}}{\partial \mathbf{R}_\nu} \, dt
+ \sum_\nu \frac{\partial\cdot \mathbf{D}_{\mu\nu}}{\partial{\bf R}_\nu}\,dt
  + d\tilde{\bf R}_\mu,
\end{align}
where        the         potential        of         mean        force
$V^{\rm  MF}({\bf  R}_1,\dots,{\bf  R}_N)$  and  the  mobility  tensor
$\mathbf{D}_{\mu\nu}({\bf  R}_1,\dots,{\bf  R}_N)$  are,  in  general,
many-body functions  of all  particle positions. The  noise increments
satisfy the fluctuation--dissipation theorem in the form
\begin{align}
  \label{FDT}
  d\tilde{\bf R}_\mu\, d\tilde{\bf R}_\nu^{T} = 2\,\mathbf{D}_{\mu\nu}\,dt,
\end{align}
which requires that $\mathbf{D}$ be symmetric positive definite.

For simplicity, we assume the Brownian particles do not interact reversibly
with each other, but they are tethered to the origin by a spring. In this way,
the potential of mean force is
\begin{align}
  V^{MF}({\bf R}_1,\cdots,{\bf R}_N)=\frac{\kappa}{2}\sum_{\nu=1}^N{\bf R}_\nu^2.
  \label{VMF-kappa}
\end{align}
This potential prevents particles  from spreading to arbitrarily large
distances.  The mobility  matrix depends on the full  set of positions
${\bf R}_1,\cdots,{\bf R}_N$  and is therefore a  many-body object. To
simplify  matters,  we approximate  it  by  pairwise contributions  of
Rotne--Prager--Yamakawa           (RPY)            form
\begin{align} \label{RPY-tensor}
&\mathbf{D}_{\mu\nu} = D_0\times
  \nonumber\\
  &\begin{cases}
\,\mathds{1}, & \mu=\nu, \\[2mm]
\frac{3 a}{4 R_{\mu\nu}}
\left[
\left(1 + \frac{2a^2}{3R_{\mu\nu}^2}\right)\mathds{1}
+ \left(1 - \frac{2a^2}{R_{\mu\nu}^2}\right){\bf e}_{\mu\nu}{\bf e}_{\mu\nu}^T
\right], & R_{\mu\nu} \ge 2a, \\[2mm]
\left[
\left( 1 - \frac{9}{32}\frac{R_{\mu\nu}}{a} \right)\mathds{1}
+ \frac{3}{32}\frac{R_{\mu\nu}}{a}{\bf e}_{\mu\nu}{\bf e}_{\mu\nu}^T
\right], & R_{\mu\nu} < 2a.
\end{cases}
\end{align}
where         $R_{\mu\nu}=|{\bf          R}_\mu-{\bf         R}_\nu|$,
${\bf  e}_{\mu\nu}=({\bf  R}_\mu-{\bf  R}_\nu)/R_{\mu\nu}$,  and  Here
$D_0 = \frac{k_BT}{6\pi\eta a}$, where  $a$ is the hydrodynamic radius
of the particles, and $\eta$ is the solvent viscosity. Because the RPY
tensor is positive definite, a standard approach to generate the noise
is to  factorize $\mathbf{D}$ via a  Cholesky decomposition: determine
the   lower   triangular   real    matrix   $\mathbf{L}$   such   that
$\mathbf{D}=\mathbf{L}\mathbf{L}^T$. Then, given independently sampled
standard  normal  variables  collected   in  $\mathbf{G}$,  the  noise
increment is
\begin{align}
  d\tilde{\bf R} = \sqrt{2\,\Delta t}\;\mathbf{L}\cdot\mathbf{G}.
\end{align}

In summary, equation (\ref{SDE-HI}) takes the form
\begin{align}
  \label{SDE-HI-Final}
  d{\bf R}_\mu
  &=
  -\frac{1}{l^2}\sum_\nu\mathbf{D}_{\mu\nu}\, \mathbf{R}_\nu \, dt
  + \sum_\nu \frac{\partial\cdot \mathbf{D}_{\mu\nu}}{\partial{\bf R}_\nu}\,dt
  + d\tilde{\bf R}_\mu,
\end{align}
with  $l  =  \sqrt{\frac{k_B   T}{\kappa}}$  being  the  trap  length.
Although the  RPY mobility is  divergence-free and the second  term on
the  right-hand side  vanishes, we  retain  it for  generality as  the
parameterized    model   introduced    below   is    not   necessarily
divergence-free.

\subsection{Parameterized model}
We   generate   trajectories   $\{{\bf   R}_\mu\}$   using   the   SDE
(\ref{SDE-HI}) with the RPY mobility (\ref{RPY-tensor}). We then treat
these  data as  if their  microscopic origin  were unknown  and assume
instead that they arise from a  coarse-grained model of the same form,
but with an unknown parameterized  mobility tensor to be inferred. For
this   parametrization   we   assume   that  the   mobility   is   (i)
translationally   invariant,  (ii)   isotropic,  and   (iii)  pairwise
additive.  Under these assumptions, the most general two-body mobility
between  distinct  particles  $\mu\neq\nu$  can  depend  only  on  the
distance $R_{\mu\nu}$ and must  be equivariant under global rotations.
For   a   rank-two   tensor   built  from   a   single   unit   vector
${\bf   e}_{\mu\nu}$,  the   only  rotationally   covariant  tensorial
structure is of the form
\begin{equation}
\overline{\mathbf D}_{\mu\nu}(R_{\mu\nu})
= A(R_{\mu\nu})\,\mathds{1} + B(R_{\mu\nu})\,{\bf e}_{\mu\nu}{\bf e}_{\mu\nu}^T,
\qquad (\mu\neq\nu),
\end{equation}
for  two  scalar functions  $A(r)$  and  $B(r)$.  We  represent  these
unknown scalar  functions in  a flexible  way by  expanding them  in a
compactly  supported  B-spline  basis.   This  leads  to  the  generic
functional form
\begin{align}
\label{RPY-params}
\overline{\mathbf{D}}_{\mu\nu} =
\begin{cases}
\tilde{D}_0\,\mathds{1}, & \mu=\nu, \\[1mm]
  \displaystyle  D_1(R_{\mu\nu})\mathds{1} + D_{\parallel}(R_{\mu\nu})
  {\bf e}_{\mu\nu}{\bf e}_{\mu\nu}^T, & \mu \neq \nu,
\end{cases}
\end{align}
where
\begin{align}
  D_1(R_{\mu\nu})
  &\equiv\sum_\alpha a_\alpha N_{\alpha,p}(R_{\mu\nu})
    \nonumber\\
  D_{\parallel}(R_{\mu\nu})
  &\equiv\sum_\alpha b_\alpha N_{\alpha,p}(R_{\mu\nu})
\end{align}
Here   $a_\alpha$,  $b_\alpha$,   and  $\tilde{D}_0$   are  adjustable
coefficients, and $\{N_{\alpha,p}(r)\}$  are normalized B-spline basis
functions    given in Appendix \ref{App:Splines}.

As  shown  in  Appendix \ref{App:NI},  the  divergence  of
$\overline{\mathbf{D}}$ is not zero  in general, and the corresponding
noise-induced drift reads
\begin{align}
&   \sum_\nu \frac{\partial}{\partial{\bf R}_\nu}\cdot \overline{\mathbf{D}}_{\mu\nu}=
  \nonumber\\
  &
  -\sum_{\nu\neq \mu} \sum_\alpha
  \left[
    (a_\alpha + b_\alpha)\, N'_{\alpha,p}(R_{\mu\nu})
    + \frac{2b_\alpha}{R_{\mu\nu}}\, N_{\alpha,p}(R_{\mu\nu})
  \right]{\bf e}_{\mu\nu}.
    \label{Noise-induced}
\end{align}

\subsection{Parameter dynamics}

To  estimate  the  dynamic   parameters  $a_\alpha$,  $b_\alpha$,  and
$\tilde{D}_0$, we couple the SDE (\ref{SDE-HI}) with the parameterized
model (\ref{RPY-params})  via evolution equations for  the parameters,
following the strategy in (\ref{dalg}).  We propose

\begin{align}
\frac{d\tilde{D}_0}{dt} &= \frac{1}{T_0}\left[ \langle O(t,{\Delta t}) \rangle_* - O(t,{\Delta t}) \right] \nonumber\\
\frac{da_\alpha}{dt} &= \frac{1}{T_\alpha}\left[ \langle G_\alpha(t,{\Delta t}) \rangle_* - G_\alpha(t,{\Delta t}) \right]\nonumber\\
\frac{db_\alpha}{dt} &= \frac{1}{T_\alpha}\left[ \langle H_\alpha(t,{\Delta t}) \rangle_* - H_\alpha(t,{\Delta t}) \right],
\label{params-evol}
\end{align}
where the observables $O$, $G_\alpha$, and $H_\alpha$ are
\begin{align}
  O(t,{\Delta t})
  &= \frac{1}{N}\sum_{\mu=1}^N \left| \Delta \mathbf{R}_{\mu} (t,{\Delta t}) \right|^2,
  \label{Ottau}
\\
  G_\alpha(t,{\Delta t})
  &= \sum_{\mu\neq\nu}
  \frac{N_{\alpha,p}(R_{\mu\nu}(t))}{\mathcal{N}_\alpha(t)}\,
  \Delta \mathbf{R}_{\mu} (t,{\Delta t})^{\niceT}\cdot
  \Delta \mathbf{R}_{\nu} (t,{\Delta t}),
  \label{Galpha}
\\
  H_\alpha(t,{\Delta t})
  &=  \sum_{\mu\neq\nu}
  \frac{N_{\alpha,p}(R_{\mu\nu}(t))}{\mathcal{N}_\alpha(t)}\Delta R^{\parallel}_{\mu} (t,{\Delta t})\Delta R^{\parallel}_{\mu} (t,{\Delta t})
  \label{Halpha}
\end{align}
where
\begin{align}
  \Delta \mathbf{R}_{\mu} (t,{\Delta t})
  &= \mathbf{R}_\mu(t-{\Delta t}) - \mathbf{R}_\mu(t)  
    \nonumber\\
  \Delta R^{\parallel}_{\mu} (t,{\Delta t})
&=  {\bf e}_{\mu\nu}^\niceT(t)\esc   \Delta \mathbf{R}_{\mu} (t,{\Delta t})
  \end{align}
and the normalization is
\begin{equation}
  \mathcal{N}_\alpha(t) \equiv \sum_{\mu\neq\nu} N_{\alpha,p}(R_{\mu\nu}(t)).
\end{equation}

\begin{figure*}[!t]
  \centering
  % --- Row 1 ---
  \begin{minipage}[b]{0.49\textwidth}
    \centering
    \includegraphics[width=\linewidth]{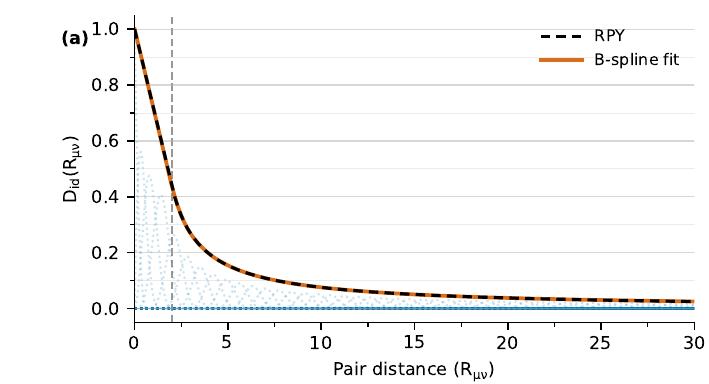}
    % If your image doesn't have the (a) label inside it, uncomment below:
    % \caption*{(a)} 
  \end{minipage}
   \hfill
    \begin{minipage}[b]{0.49\textwidth}
    \centering
    \includegraphics[width=\linewidth]{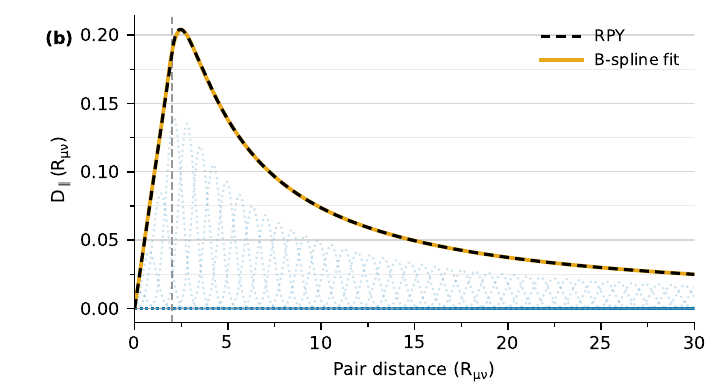}
  \end{minipage}
% \begin{minipage}[b]{0.48\textwidth}
  %   \centering
  %   \includegraphics[width=\linewidth]{Fig1.pdfrpy_bspline_vs_theory_tau.pdf}
  % \end{minipage}
  
  \vspace{-1ex} % Pull caption closer to images
  \caption{Validation of the  parameterized mobility model.  {Vertical
      dashed line corresponds to $R_{\mu\nu} = 2a$.}  {\bf (a)} RPY vs
    learned B-spline reconstruction of $D_1(R_{\mu\nu})$. Dotted lines
    show  the   spline  basis  functions  multiplied   by  the  fitted
    parameters     $a_\alpha$.      {\bf    (b)}     Similarly     for
    $D_{\parallel}(R_{\mu\nu})$.  }
  \label{fig:rpy_validation_2x2}
\end{figure*}

The choice of these specific coarse-grained observables is dictated by
the  structure  of  the parameterized  model  (\ref{RPY-params}).   In
particular, the time-dependence of $\tilde{D}_0$, which multiplies the
self-mobility ($\mu=\nu$), is naturally  associated with a mean-square
displacement,  motivating  the  definition  of  $O(t,{\Delta  t})$  in
(\ref{Ottau}). The  coefficients $a_\alpha$ govern the  isotropic part
of the  pair mobility at  separations selected by the  basis functions
$N_{\alpha,p}$,   motivating   the  distance   resolved   displacement
observable  $G_\alpha(t,{\Delta  t})$ in  (\ref{Galpha}).   Similarly,
$b_\alpha$     controls     the    anisotropic     component     along
${\bf     e}_{\mu\nu}$,      motivating     the      definition     of
$H_\alpha(t,{\Delta t})$ in (\ref{Halpha}).

The evolution  equations (\ref{params-evol})  enforce the  matching of
time--lagged displacement correlations at a fixed lag ${\Delta t}$. As
discussed  in Appendix  \ref{App:sign}, the  sign in  the updates  for
$\tilde D_0$,  $a_\alpha$, and $b_\alpha$  is dictated by the  sign of
the  response  of  the  targeted   correlation  with  respect  to  the
parameter.

In contrast to the analytical  treatment of a single Brownian particle
in  Section~\ref{B-1D},  the  system  defined  by  (\ref{SDE-HI})  and
(\ref{RPY-tensor}) admits  no closed-form solution.   Consequently, we
obtain  the  ``ground-truth'' averages  $\llangle\cdots\rrangle_*$  in
(\ref{params-evol}) by numerically integrating (\ref{SDE-HI}) with the
RPY tensor (\ref{RPY-tensor}) with  given parameters. We then simulate
the    SDE   (\ref{SDE-HI})    using   the    parameterized   mobility
(\ref{RPY-params}),   simultaneously  evolving   its  parameters   via
(\ref{params-evol}). Under this coupled dynamics, the system converges
to a fixed point in parameter space,  which we take as our estimate of
the dynamic parameters in the coarse-grained model.

\subsection{Results and discussion}
We  have  simulated  the coupled  equations  (\ref{SDE-HI-Final})  and
(\ref{params-evol})      using       the      parameterized      model
${\bf     D}_{\mu\nu}=\overline{\bf     D}_{\mu\nu}$    defined     in
(\ref{RPY-params}).    The   simulation   details  are   reported   in
Appendix~\ref{App:sim-details}.   The  parameter  dynamics   converges  towards  a  fixed  point
$\tilde{D}_0^*,a_\alpha^*,b_\alpha^*$.

Figure~\ref{fig:rpy_validation_2x2}  compares   the  learned  B-spline
reconstruction   with  the   analytical   RPY   tensor.  The   learned
representation accurately  reproduces the analytical result:  both the
identity  component and  the  parallel (dyadic)  component follow  the
theoretical  curves   with  no   visible  systematic   deviations.  In
particular,  the reconstruction  correctly  captures the  near-contact
behavior around  $R_{\mu\nu}=2a$ as  well as  the long-range  decay at
larger separations.

A practical limitation appears only  near the boundaries of the fitted
range. Since the  parameter updates are performed at  every time step,
the effective statistics contributing to a given coefficient depend on
how often particle pairs sample  separations within the support of the
corresponding B-spline basis function.  Near $R_{\mu\nu}\approx 0$ and
close to the cutoff, the number of contributing pairs is significantly
reduced, leading  to noisier  updates. Consequently,  the coefficients
associated  with  basis  functions  supported near  the  ends  of  the
interval  converge  more  slowly   than  those  corresponding  to  the
well-sampled central region.

\begin{figure}[!t]
  \begin{minipage}[b]{0.48\textwidth}
    \centering
    \includegraphics[width=\linewidth]{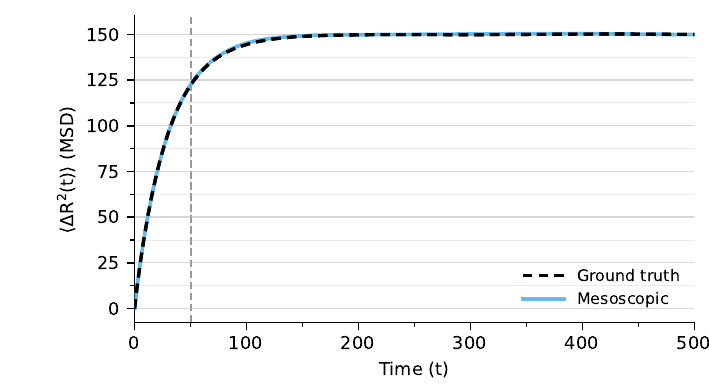}
  \end{minipage}

  \vspace{-1ex} % Pull caption closer to images
  \caption{Validation   of    the   parameterized    mobility   model.
    Mean-squared  displacement: ground  truth (black  dashed line)  vs
    mesoscopic model (blue solid line). The vertical line is at $t=50$
    and signals  the range of  lag times  $\Delta t<50$ for  which the
    parameters do not depend on $\Delta t$.}
  \label{fig:RMS}
\end{figure}

Because  the RPY  mobility tensor  is reproduced  faithfully with  the
B-splines  parameterized  model,  all observables  computed  with  the
original RPY model will be reproduced with the parameterized model. As
an example, we plot in  Fig \ref{fig:RMS} the mean square displacement
in both  models, showing  excellent agreement.   As the  particles are
tethered to  the origin by a  spring, the RMS displacement  exhibits a
plateau.  This  plot  helps  to  understand the  range  of  time  lags
$\Delta  t$ for  which  the converged  parameters  are insensitive  to
$\Delta t$. In the plateau region  $\Delta t\ge 50$ the observable $O$
in  (\ref{Ottau}) does  not  allow one  to  discriminate well  between
different parameters,  leading to  a dependence  of the  parameters on
$\Delta  t$. We  observe that  the lag  time range  for the  parameter
$\tilde{D}_0$ is  ${\Delta t} \in \left[0.1,50\right]$,  while for the
off-diagonal      parameters       $a_\alpha,b_\alpha$      it      is
${\Delta t}  \in \left[0.1,2\right]$.  In the present work, for simplicity, we use the same update frequency,
relaxation time, and lag time for all basis coefficients. This is only
an implementation choice, not a restriction of the method. In general,
each parameter could be assigned its own update frequency, relaxation
time, and lag time. One could also use adaptive updates based on the
number of effective samples contributing to each basis coefficient.
Such non-uniform updates would act as a preconditioning of the
parameter dynamics. They do not, however, generate additional
information in poorly sampled regions; improving those regions would
require longer simulations, enhanced sampling, or appropriately
reweighted estimators. More generally, the self-averaging formulation
is modular: different parameters may be coupled to different
instantaneous estimators and different update protocols, as long as the
corresponding averaged dynamics has the desired moment-matching
condition as its fixed point. In this work we kept the same conditions
for all parameters in order to keep the implementation and the analysis
of lag dependence as simple as possible.

\newpage\section{Example 3: LJ binary mixture}
\label{Sec:MD}
Once the self-averaging method has  been validated in the previous two
examples—where data  are generated from  a known mesoscopic  model and
the parameters of a flexible  coarse-grained model are recovered so as
to reproduce  the original  dynamics—we apply  it to  infer nontrivial
physics in a molecular system.

We consider a LJ  binary mixture in a cubic periodic  box of side $L$,
containing $N_A$ light particles (species  $A$) and $N_B$ heavy tracer
particles  (species  $B$). All  pairs  interact  through the  same  LJ
potential,
\begin{equation}
u_{\alpha\beta}(r)=4\varepsilon\Big[\big(\sigma/r\big)^{12}-\big(\sigma/r\big)^{6}\Big],
\qquad
\sigma=1,\ \varepsilon=1,
\end{equation}
so that the only distinction between the species is inertia:
\begin{equation}
m_B = C\, m_A,\qquad C\gg 1.
\end{equation}
The  number   density  of  particles  is   $\rho=(N_A+N_B)/L^3$.   We
equilibrate  the full  $A\!+\!B$  system using  Langevin-thermostatted
dynamics at fixed temperature $T$.  After equilibration, we remove the
thermostat  and   perform  an   NVE  production  run,   recording  the
heavy-particle      positions      and      velocity      trajectories
$\{\mathbf R_\mu(t),\mathbf V_\mu(t)\}_{\mu=1}^{N_B}$.

\subsection{CG model}
We model  the dynamics of the  heavy particles with a  set of Langevin
equations of the form
\begin{align}
d\mathbf R_\mu &= \mathbf V_\mu\,dt,
\nonumber\\
d\mathbf P_\mu &=
-\nabla_{\mathbf R_\mu}V^{\rm MF}(\mathbf R)\,dt
-\sum_{\nu}\boldsymbol{\Gamma}_{\mu\nu}(\mathbf R)\esc\mathbf V_\nu\,dt
                 \nonumber\\
  &+\sum_{\nu}\Sigma_{\mu\nu}(\mathbf R)\,d\mathbf W_\nu,
\label{LJ-SDE}
\end{align}
where $\mathbf P_\mu = m_B \mathbf V_\mu$ denotes the momentum of the
$\mu$-th heavy particle. The potential of mean force is approximated
using the standard pairwise form 
\begin{equation}
V^{\rm MF}(\mathbf R_1,\ldots,\mathbf R_{N_B})
= \sum_{\mu<\nu} U(R_{\mu\nu}),
\label{eq:VMF_LJ}
\end{equation}
which is expected  to be accurate at dilute  tracer concentrations and
is  sufficient  for  the  present  validation  benchmarks.   The  pair
potential  $U(r)$  is  represented  in a  compactly  supported  radial
B-spline basis of degree $p$ on $[0,r_c]$,
\begin{equation}
U(r)=\sum_{i=0}^{M-1}\lambda_i\,N_{i,p}(r),
\label{eq:U_spline}
\end{equation}
with clamped knot vector $\{t_0,\ldots,t_{M+p}\}$.
  All   pair   distances    $R_{\mu\nu}$   and   unit   vectors
$\mathbf e_{\mu\nu}$  are computed using the  minimum-image convention
under periodic boundary conditions.

The friction matrix is defined as the inverse of the mobility matrix
\begin{equation}
\boldsymbol{\Gamma}(\mathbf R)=\boldsymbol{\mu}(\mathbf R)^{-1}.
\label{eq:Gamma_def}
\end{equation}
We postulate a configuration-dependent mobility matrix with translationally
invariant, isotropic, pairwise structure,
\begin{align}
\boldsymbol{\mu}_{\mu\nu}(\mathbf R) &=
\begin{cases}
\mu_0\,\mathds{1}, & \mu=\nu,\\[2pt]
\displaystyle \mu_\perp(R_{\mu\nu})\,\mathbf P_\perp(\mathbf e_{\mu\nu})
+\mu_\parallel(R_{\mu\nu})\,\mathbf P_\parallel(\mathbf e_{\mu\nu})
, & \mu\neq \nu,
\end{cases}
\label{eq:mu_ansatz_LJ}
\end{align}
where
\begin{align}
\mu_\perp(R_{\mu\nu})&\equiv  \sum_{\alpha=0}^{M-1}N_{\alpha,p}(R_{\mu\nu})a_\alpha
           \nonumber\\
\mu_\parallel(R_{\mu\nu})&\equiv  \sum_{\alpha=0}^{M-1}N_{\alpha,p}(R_{\mu\nu})b_\alpha
\end{align}
and the two projectors are
\begin{align}
  \mathbf P_\parallel({\bf e}_{\mu\nu})
  &=\mathbf e_{\mu\nu}\mathbf e_{\mu\nu}^{\mathsf T}
    \nonumber\\
  \mathbf P_\perp({\bf e}_{\mu\nu})
  &=\mathds{1}-\mathbf P_\parallel({\bf e}_{\mu\nu})  
\end{align}
where the unit vector is $\mathbf e_{\mu\nu}=(\mathbf R_\mu-\mathbf R_\nu)/R_{\mu\nu}$.

We also tested an alternative irreversible model in which the friction
matrix  $\boldsymbol{\Gamma}(\mathbf  R)$ was  parameterized  directly
using a  pairwise B-spline  ansatz. This choice  was motivated  by the
work  of Bocquet  \emph{et al.}~\cite{bocquet1997},  who analyzed  the
Langevin  dynamics of  two infinitely  massive particles  in terms  of
pair-distance–dependent friction coefficients obtained from Green–Kubo
relations, with  appropriate corrections for finite-size  effects.  In
our simulations,  however, the  parameters associated with  a pairwise
distance-dependent  friction tensor  did  not  converge reliably.   In
contrast,   the   pairwise  mobility   model   (\ref{eq:mu_ansatz_LJ})
consistently exhibited stable relaxation  of the parameters. A natural
interpretation  is  that  the   matrix  inverse  introduces  effective
many-body contributions in the friction,  which appear to be important
in  the  present LJ  system  for  reproducing the  tracer  correlation
targets, whereas  enforcing a  strictly pairwise  friction may  be too
restrictive in this system.   This empirical observation motivates the
mobility parametrization used below.

In the noise  term of the SDE (\ref{LJ-SDE}), $\mathbf  W$ is a $3N_B$
dimensional   Wiener    process   and   the   noise    satisfies   the
fluctuation--dissipation relation
\begin{equation}
\boldsymbol{\Sigma}(\mathbf R)\boldsymbol{\Sigma}(\mathbf R)^{\mathsf T}
=2k_BT\,\boldsymbol{\Gamma}(\mathbf R).
\label{eq:FDT_Gamma}
\end{equation}

The   model~(\ref{LJ-SDE})    with   pairwise   mobility    given   by
(\ref{eq:mu_ansatz_LJ}) will be referred  to as the \textit{HI model}.
We compare  it with  a simplified  \textit{Stokes model},  obtained by
setting      $a_\alpha=b_\alpha=0$.       This     choice      implies
$\mu_\parallel=\mu_\perp=0$,    thereby    eliminating    hydrodynamic
interactions   between the heavy  particles.  In this  limit, only
the  Stokes friction  exerted by  the implicit  solvent remains.   The
comparison between these two models  allows us to isolate and quantify
the role of hydrodynamic interactions in the system.

\subsection{Parameter dynamics}

We estimate  the conservative parameters $\{\lambda_i\}$  and mobility
parameters $\mu_0,\{a_\alpha,b_\alpha\}$  by coupling the  CG dynamics
to  feedback laws  that  relax instantaneous  estimators toward  their
ground-truth expectations:
\begin{align}
\frac{d\lambda_i}{dt} &= -\frac{1}{T_\lambda}\Big[\langle W_i\rangle_* - W_i(t)\Big],
\nonumber\\
\frac{d\mu_0}{dt} &= \frac{1}{T_0}\Big[\langle O(\tau_S)\rangle_* - O(t,\tau_S)\Big],
\nonumber\\
\frac{da_\alpha}{dt} &= -\frac{1}{T_\alpha}\Big[\langle G^{\perp}_\alpha(\tau_{\perp})\rangle_* - G^{\perp}_\alpha(t,\tau_{\perp})\Big],
\nonumber\\
\frac{db_\alpha}{dt} &= -\frac{1}{T_\alpha}\Big[\langle G^{\parallel}_\alpha(\tau_{\parallel})\rangle_* - G^{\parallel}_\alpha(t,\tau_{\parallel})\Big]
\label{eq:evol_b}
\end{align}
\begin{figure*}[t]
  \centering
  \begin{subfigure}[t]{0.49\linewidth}
    \includegraphics[width=\linewidth]{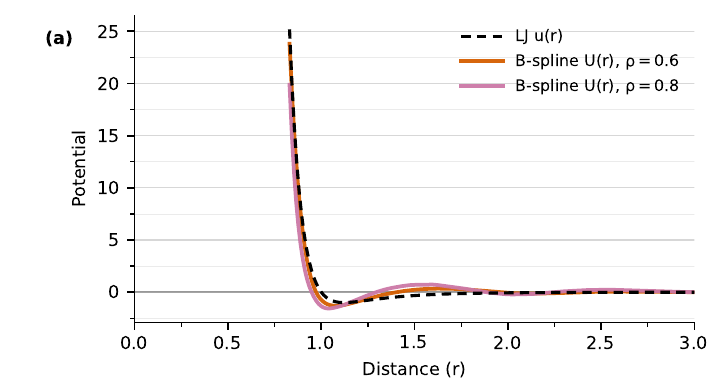}
 %   \caption{$V^{\rm MF}(r), V^{\rm LJ}(r)$}
  \end{subfigure}\hfill
  \begin{subfigure}[t]{0.49\linewidth}
    \includegraphics[width=\linewidth]{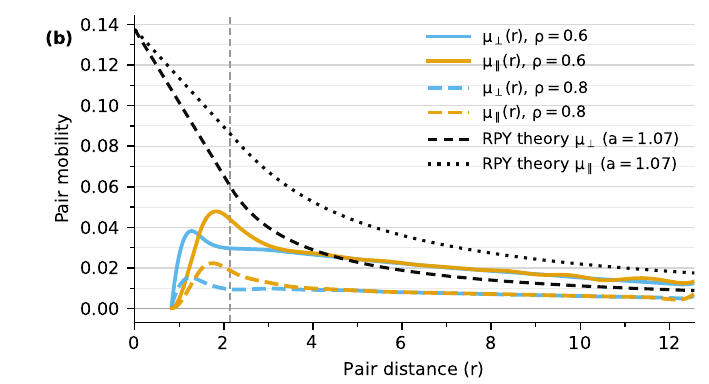}
%    \caption{Reconstructed $\mu_\perp(r)$ and $\mu_\parallel(r)$ }
  \end{subfigure}
  \caption{Fitted  building blocks  at  two  solvent number  densities
    $\rho=0.6,0.8$  and  temperature  $T=2$.   {\bf  (a)}  Fitted
    potential of  mean force,  compared with  the LJ  potential.  {\bf
      (b)} Transverse and longitudinal  mobilities.  The vertical line
    is at the  hydrodynamic radius for which the  RPY changes behavior
    by construction. }
  \label{Fig:LJ-1}
\end{figure*}
with observables
\begin{align}
W_i(t)
&=\frac{1}{N_B(N_B-1)}\sum_{\mu<\nu} N_{i,p}(R_{\mu\nu}(t)),
\nonumber\\
O(t,\tau_S)
&=\frac{1}{N_B}\sum_{\mu=1}^{N_B}\left| \Delta \mathbf{R}_{\mu} (t,\tau_S) \right|^2,
\nonumber\\
G^\perp_\alpha(t,\tau_\perp)
&=\sum_{\mu<\nu} \frac{N_{\alpha,p}(R_{\mu\nu}(t))}{\mathcal N_\alpha(t)}\,
  {\bf V}_{\mu\nu}^{\perp\niceT}(t)\esc {\bf V}_{\mu\nu}^\perp(t-\tau_\perp)
  \nonumber\\
G^\parallel_\alpha(t,\tau_\parallel)
&=\sum_{\mu<\nu} \frac{N_{\alpha,p}(R_{\mu\nu}(t))}{\mathcal N_\alpha(t)}\,
  {\bf V}_{\mu\nu}^{\parallel\niceT}(t)\esc {\bf V}_{\mu\nu}^\parallel(t-\tau_\parallel)
\label{LJ-observables}
\end{align}
where
\begin{align}
  {\bf V}_{\mu\nu}^\perp(t)
  &=\mathbf P_\perp(\mathbf e_{\mu\nu}(t))\esc {\bf V}_{\mu\nu}(t)
\nonumber\\
  {\bf V}_{\mu\nu}^\perp(t-\tau_\perp)
    &=\mathbf P_\perp(\mathbf e_{\mu\nu}(t))\esc {\bf V}_{\mu\nu}(t-\tau_\perp)
      \nonumber\\
  {\bf V}_{\mu\nu}^\parallel(t)
  &=\mathbf P_\parallel(\mathbf e_{\mu\nu}(t))\esc {\bf V}_{\mu\nu}(t)
\nonumber\\
  {\bf V}_{\mu\nu}^\parallel(t-\tau_\parallel)
    &=\mathbf P_\parallel(\mathbf e_{\mu\nu}(t))\esc {\bf V}_{\mu\nu}(t-\tau_\parallel)
\end{align}
The                relative                 velocity                is
$\mathbf V_{\mu\nu}=\mathbf V_\mu-\mathbf V_\nu$ and the normalization
is $\mathcal  N_\alpha(t)=\sum_{\mu<\nu} N_{\alpha,p}(R_{\mu\nu}(t))$.
Here
$\Delta\mathbf    R_\mu(t,\tau_S)\equiv    \mathbf    R_\mu(t)-\mathbf
R_\mu(t-\tau_S)$  denotes a  lagged  displacement increment  (computed
using unwrapped tracer trajectories under PBC).

The static observable $W_i$ is  a distance-resolved occupancy that, in
the limit  of fine binning,  is equivalent to the  radial distribution
functions (RDF) $g_{BB}(r)$ of  heavy particles.  With this selection,
we aim  at matching  RDF.  The  dynamical observable  $O(t,\tau_S)$ is
chosen  as  the mean-squared  displacement  (MSD),  which targets  the
self-mobility scale $\mu_0$.

Finally, the observables $G^\parallel_\alpha$ and $G^\perp_\alpha$ are
relative-velocity   time  auto-correlations   evaluated  at   fixed   lags
$\tau_\parallel$  and   $\tau_\perp$  and  fixed   relative  distance,
designed to target the transverse  and longitudinal pair components of
the    mobility    parametrization    \eqref{eq:mu_ansatz_LJ}.     The
distance-resolved  observables  defined  in  (\ref{LJ-observables})  are
constructed  to  respect  the  symmetries  of  the
dynamics:  they  are  translationally  invariant  (depending  only  on
separations and  velocity differences), Galilean  invariant (invariant
under  uniform  boosts),  and rotationally  invariant  through  scalar
bilinear    forms     built    with    the     covariant    projectors
$\mathbf  P_{\parallel/\perp}$.    Each  pair  contribution   is  also
symmetric  under  exchange  $\mu\leftrightarrow\nu$,  since  the  sign
changes    $\mathbf    e_{\mu\nu}\!\to\!-\mathbf    e_{\mu\nu}$    and
$\mathbf V_{\mu\nu}\!\to\!-\mathbf V_{\mu\nu}$  cancel in the bilinear
forms.  Finally,  the normalization by $\mathcal  N_\alpha(t)$ reduces
sensitivity to fluctuations in the number of pairs contributing within
the support of a given spline at a given time.

\subsection{Results and discussion}
All-atom MD simulations of the  LJ binary mixture were performed using
LAMMPS \cite{thompson2022}  while the  CG dynamics  (\ref{LJ-SDE}) was
integrated  with  the  G-JF  algorithm, as  in  Sec.~\ref{B-1D}.   The
details    of    the    simulations     are    given    in    appendix
\ref{App:sim-details}.  Plots shown are for a mass ratio of $10$.

\subsubsection{The parameterized CG model}

\begin{figure}[h]
  \includegraphics[width=\linewidth]{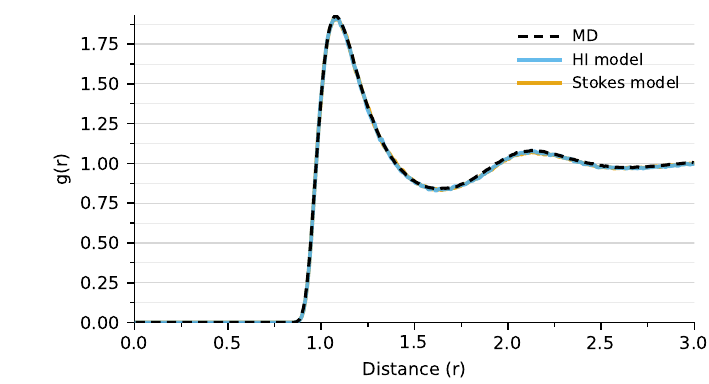}
  \caption{Radial distribution function $g_{BB}(r)$ computed from three distinct models, showing excellent mutual agreement.}
  \label{Fig:RDF}
\end{figure}

In Fig ~\ref{Fig:LJ-1}  we show the pair-wise potential  of mean force
(panel  (a))  and  the  pair mobilities (panel  (b))  that  are
obtained  after  convergence  of   the  evolution  of  the  parameters
according   to   the   coupled   evolution   of   (\ref{LJ-SDE})   and
(\ref{eq:evol_b}).   These are  the building  blocks of  the CG  model
(\ref{LJ-SDE})  describing  the  dynamics   of  the  heavy  particles.
Observe in  Fig \ref{Fig:RDF}  that the RDF  $g_{BB}(r)$ of  the heavy
particles is essentially zero below a hard-core distance.

In our  data, the first nonzero  RDF bin occurs at  $r\simeq 0.83$, so
shorter separations  are never sampled  and any basis  supported there
would   lead    to   near-zero   occupancies.    We    therefore   set
$r_{\min}=0.83$, as the left endpoint of all radial spline expansions.
In  Fig \ref{Fig:LJ-1},  panel (a)  shows that  the potential  of mean
force is  slightly different  from LJ  potential, indicating  that 
    the light-particle bath (species~A) induces a weak solvation effect on
    the heavy tracers (species~B) that grows with the number density of the
    light-particle background.%
  In panel  (b) we  plot the  pair mobilities
$\mu_\parallel(R_{\mu\nu}),\mu_\perp(R_{\mu\nu})$,  which are  defined
up to  the maximum  accessible separation $L/2$.   At this  scale they
have clearly not decayed  to zero, indicating long-ranged hydrodynamic
interactions between the heavy particles.   At short distances, of the
order of the  LJ interaction range, the  mobilities display overshoots
that  reflect  strong  interaction  effects,  likely  associated  with
structural  layering  of  light   particles  aroung  heavy  ones.  The
intensity of HI decreases with increasing solvent density.

Remarkably,  for   separations  $4   \lesssim  r  \lesssim   L/2$  the
longitudinal and transverse mobilities coincide,
\begin{align}
  \mu_\parallel(r)=\mu_\perp(r)\equiv \mu(r).
  \label{mu=mu}
\end{align}
Substituting this  result into Eq.~(\ref{eq:mu_ansatz_LJ})  shows that
the pair mobility tensor becomes isotropic in this range,
\begin{align}
\boldsymbol{\mu}_{\mu\nu}({\bf R}_{\mu\nu})=\mu(R_{\mu\nu})\,\mathbb{1}.  
\end{align}
The inverse  matrix, i.e.\  the friction  matrix, therefore  takes the
block form
\begin{align}
\boldsymbol{\Gamma}_{\mu\nu}=[\mu^{-1}]_{\mu\nu}\,\mathbb{1}.  
\end{align}
As  a  consequence,  the   friction  force  in  Eq.~(\ref{LJ-SDE})  is
proportional to the velocity, with a configuration-dependent many-body
friction coefficient.  In this  sense, the heavy particles effectively
perceive  an   \textit{isotropic}  medium,   where  the   friction  is
determined  by  the  instantaneous configuration  of  the  surrounding
particles.

\begin{figure*}[t]
  \centering

  \begin{subfigure}[t]{0.49\linewidth}
    \includegraphics[width=\linewidth]{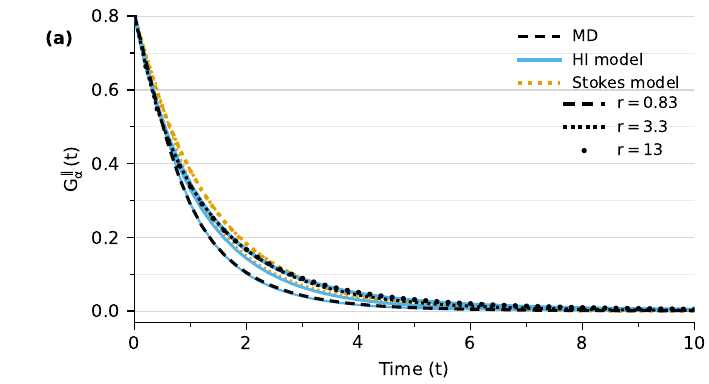}
  \end{subfigure}\hfill
  \begin{subfigure}[t]{0.49\linewidth}
    \includegraphics[width=\linewidth]{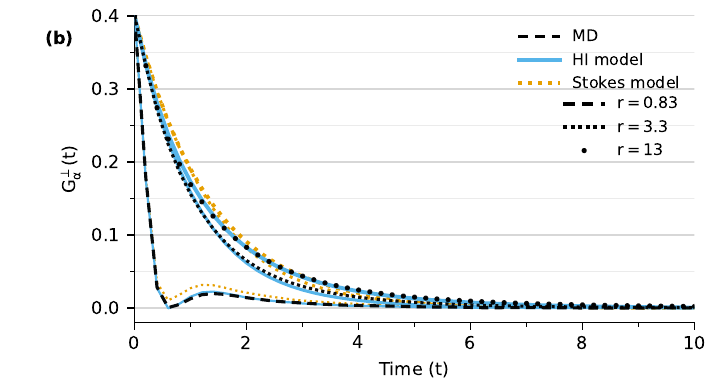}
  \end{subfigure}
  \caption{Comparison  between  the  HI  model and  the  Stokes  model
    against  MD results  for the  distance-resolved relative  velocity
    autocorrelation  functions  $G^{\parallel}_\alpha(t)$ (panel  (a))
    and    $G^{\perp}_\alpha(t)$    (panel   (b))    at    separations
    $r=\alpha\Delta  r$  ($\Delta r  =  0.59$).   The HI  model  shows
    excellent agreement with the MD data, while the Stokes model fails
    to reproduce the observed behavior.}
  \label{fig:G-H}
\end{figure*}

In Fig.~\ref{Fig:LJ-1}(b), we also  plot, for comparison, the mobility
predicted  by the  anisotropic  RPY model.   The  RPY tensor  predicts
distinct     longitudinal    and     transverse    couplings,     with
$\mu_\parallel(r)>\mu_\perp(r)$     and     an    asymptotic     ratio
$\mu_\parallel/\mu_\perp \to 2$ at  large separation.  By contrast the
LJ system displays  isotropic mobility $\mu_\parallel(r)=\mu_\perp(r)$
at  intermediate  distances. 
This indicates that, at the level of the inferred pair mobility, the LJ
tracer-bath system differs from the RPY form in this regime.
 Note that the
RPY  mobility tensor  emerges  from  a continuum,  low-Reynolds-number
approximation for rigid spherical particles of prescribed hydrodynamic
radius suspended in an incompressible viscous solvent. By contrast, in
the present LJ mixture the tracer  particles differ from the bath only
through  their inertia.  In particular,  there  is no  {\em a  priori}
geometric radius  or boundary  condition that  would justify  a direct
identification of  the tracers  with continuum  spheres.  Accordingly,
the RPY curves—constructed using $D_0$ measured from MD and an assumed
hydrodynamic  radius  $a$—  in  Fig.   \ref{Fig:LJ-1}  (b)  should  be
regarded only as a qualitative
reference for the shape of the inferred pair
mobility, not as a validation metric for the fitted CG model.

At large separations, the parallel mobility develops mild oscillations
that we attribute to finite-size effects and low pair statistics.

\newpage\subsubsection{Validation of the parameterized CG model}

To validate the  model (\ref{LJ-SDE}) with the  fitted building blocks
shown   in  Fig.~\ref{Fig:LJ-1},   we   compare  several   macroscopic
observables computed in both the microscopic MD simulations and the CG
model.   

The   first   observable   we   consider   is   the   RDF   shown   in
Fig.~\ref{Fig:RDF}.    Both  models—with   and  without   hydrodynamic
interactions—produce indistinguishable RDFs, which  are also very well
reproduced by  the CG model. The  agreement between the HI  and Stokes
models  is expected,  since the  RDF  is a  purely static  equilibrium
structural property and  therefore does not depend  on the dissipative
part of the dynamics.

Next, we consider a dynamical observable: the distance-resolved relative
velocity autocorrelations. These are defined as the time-origin averages
of the observables introduced in Eq.~(\ref{LJ-observables})
\begin{align}
  G^\perp_\alpha(\tau)
  &=\frac{1}{T}\int_0^T dtG^\perp_\alpha(t,\tau)
    \nonumber\\
  G^\parallel_\alpha(\tau)
  &=\frac{1}{T}\int_0^T dtG^\parallel_\alpha(t,\tau)
\end{align}
for sufficiently large $T$.
These averages should coincide,  by construction, at $\tau=\tau_\perp$
and  $\tau=\tau_\parallel$,  respectively,   since  the  algorithm  is
designed to enforce this condition. In Fig.~\ref{fig:G-H} we show that
the results of the  mesoscopic HI model~(\ref{LJ-SDE}) closely match
the MD data  \textit{over the entire range} of  $\tau$. This agreement
provides  strong  evidence  that  the model  accurately  captures  the
dynamics of the heavy particles.  Moreover, together with the observed
independence of the  inferred parameters from the lag  time $\Delta t$
used to match  time correlations, it further supports  the validity of
the  Markovian  description.  By   contrast,  the  Stokes  model—which
neglects hydrodynamic interactions—does not reproduce the MD data with
the same level of accuracy as the HI model.

\begin{figure}[t]
    \includegraphics[width=\linewidth]{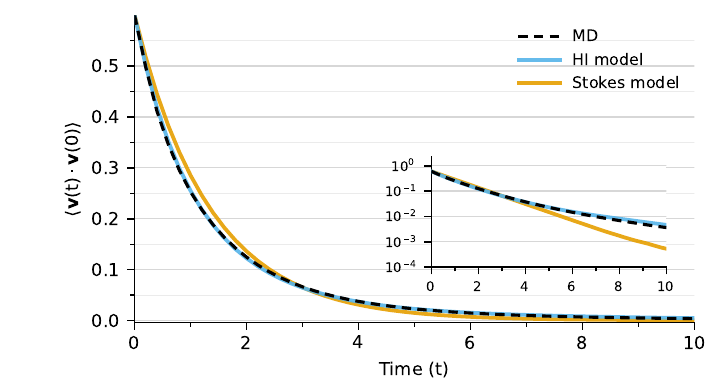}
    \caption{Velocity autocorrelation function (VACF) obtained from MD
      simulations and  from the HI  and Stokes models. The  main panel
      shows  linear–linear scales,  while  the inset  is presented  in
      linear–logarithmic   scales.   The   Stokes  model   yields   an
      exponential  decay that  is  inconsistent with  the MD  results,
      whereas the HI model reproduces the observed behavior.}
  \label{Fig:VACF}
\end{figure}
We also compare  the VACF of the heavy particles  obtained from MD and
from the CG  models. In the absence of  hydrodynamic interactions, the
VACF  decays  exponentially,  as shown  in  Fig.~\ref{Fig:VACF}.  This
behavior clearly does  not match the MD results, which  exhibit a much
slower decay.
By contrast, the CG model with hydrodynamic interactions reproduces the
MD VACF very well.

\begin{figure}[t]
  \centering

  \begin{subfigure}[t]{\linewidth}
    \includegraphics[width=\linewidth]{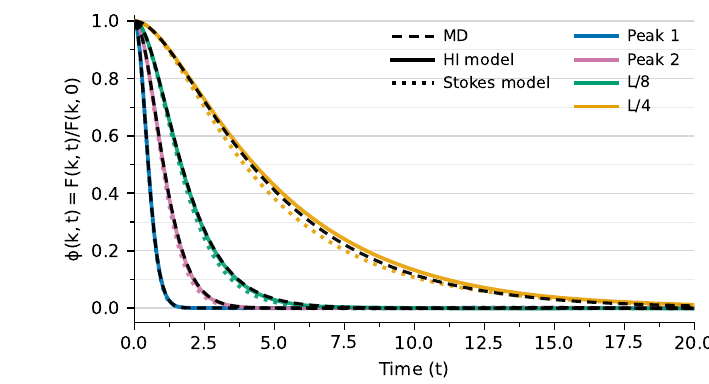}
  \end{subfigure}
  
  \begin{subfigure}[t]{\linewidth}
    \includegraphics[width=\linewidth]{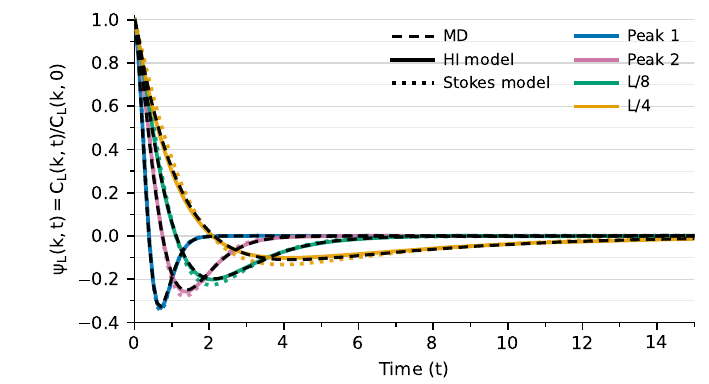}
  \end{subfigure}\hfill

  \begin{subfigure}[t]{\linewidth}
    \includegraphics[width=\linewidth]{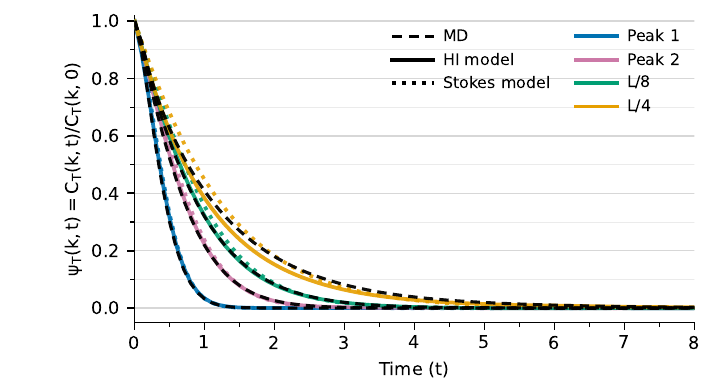}
  \end{subfigure}

\caption{Comparison of the HI and Stokes
models with MD results for the collective
correlation functions (\ref{eq:collective_norm}). The comparison is performed at
different wavelengths, corresponding to the first and second peaks of
the radial distribution function, as well as to wavelengths equal to
$L/8$ and $L/4$, where $L$ is the simulation box size.}
  \label{fig:lj_mixture_5x2}
\end{figure}

To  probe  collective  tracer  dynamics beyond  the  single  and  pair
particle diagnostics,  we measure normalized correlation  functions of
tracer density and  current modes at selected wavevectors  $k$.  For a
given reciprocal wavevector $\mathbf k$, we define the coherent tracer
density mode
\begin{equation}
\rho_{\mathbf k}(t)=\sum_{\mu=1}^{N_B} e^{i\mathbf k\cdot \mathbf R_\mu(t)},
\end{equation}
and the corresponding tracer current mode
\begin{equation}
\mathbf J_{\mathbf k}(t)=\sum_{\mu=1}^{N_B} \mathbf V_\mu(t)\,
e^{i\mathbf k\cdot \mathbf R_\mu(t)}.
\end{equation}
Writing $\hat{\mathbf k}=\mathbf k/|\mathbf k|$, we decompose the current into
longitudinal and transverse parts,
\begin{equation}
J_L(t)=\hat{\mathbf k}\cdot \mathbf J_{\mathbf k}(t),\qquad
\mathbf J_T(t)=\mathbf J_{\mathbf k}(t)-\hat{\mathbf k}\,J_L(t).
\end{equation}
We then form time-origin averaged autocorrelations,
\begin{align}
F(\mathbf k,\tau)
&= \big\langle \rho_{\mathbf k}(t+\tau)\,\rho_{\mathbf k}(t)^{*}\big\rangle,
\nonumber\\
C_T(\mathbf k,\tau)
&= \big\langle \mathbf J_T(t+\tau)\cdot \mathbf J_T(t)^{*}\big\rangle,
\nonumber\\
C_L(\mathbf k,\tau)
&= \big\langle J_L(t+\tau)\,J_L(t)^{*}\big\rangle,
\end{align}
and report the normalized shell-averaged correlators
\begin{align}
\phi(k,\tau)
&=\frac{\langle F(\mathbf k,\tau)\rangle_{\rm shell}}
        {\langle F(\mathbf k,0)\rangle_{\rm shell}},
\nonumber\\
\psi_T(k,\tau)
&=\frac{\langle C_T(\mathbf k,\tau)\rangle_{\rm shell}}
        {\langle C_T(\mathbf k,0)\rangle_{\rm shell}},
\nonumber\\
\psi_L(k,\tau)
&=\frac{\langle C_L(\mathbf k,\tau)\rangle_{\rm shell}}
        {\langle C_L(\mathbf k,0)\rangle_{\rm shell}},
\label{eq:collective_norm}
\end{align}
where  $\langle\cdot\rangle_{\rm shell}$  denotes  an  average over  a
paired  set  of $\pm\mathbf  k$  directions  with the  same  magnitude
$k=|\mathbf k|$.  In equilibrium these  quantities depend only  on $k$
and $\tau$.  The correlator $\phi(k,\tau)$ is  the normalized coherent
intermediate   scattering   function,   while   $\psi_T(k,\tau)$   and
$\psi_L(k,\tau)$  probe   transverse  (shear-like)   and  longitudinal
(compressional or sound-like) current relaxation, respectively.

We evaluate  the collective correlators  (\ref{eq:collective_norm}) at
several  wavevectors $k  =  5.79,2.96,1.96,0.98$ (in  LJ  units).  The  two
largest  wavevectors  probe  short   wavelengths,  comparable  to  the
characteristic length  scales associated  with the  first ($R_1=1.08$)
and  second ($R_1=2.12$)  peaks of  the radial  distribution function.
The  remaining  two correspond  to  wavelengths  $L/8=3.2$  and
$L/4=6.4$,  respectively, where  the  influence  of periodic  boundary
conditions remains moderate but begins  to be noticeable.  As shown in
Fig.   \ref{fig:lj_mixture_5x2}, the  HI model  provides a  systematic
improvement  over  the Stokes  model  at  the wavevectors  considered,
especially at the shortest length scales.

Wavevectors smaller (i.e.,  length scales larger) than  those shown in
Fig.~\ref{fig:lj_mixture_5x2}  are strongly  affected by  periodicity,
indicating  that  the  model  does  not  reliably  capture  collective
behavior at  scales approaching  the system  size. This  limitation is
consistent  with the  pair mobility  reported in  Fig.~\ref{Fig:LJ-1} (b),
which remains  long-ranged and  non-negligible even at  separations of
order $L/2$.

Note that, due to the  minimum image convention, the effective support
of the mobility  is restricted to distances up to  $L/2$, and its form
beyond  this range  is not  defined.  As a  consequence, a  consistent
periodic summation—such as Ewald-type methods, which require knowledge
of  the mobility  at all  distances—cannot be  constructed within  the
present coarse-grained framework. Extending  the support would require
larger  system  sizes and  increased  computational  cost, which  lies
beyond the scope of the present work and remains an open problem.

\subsubsection{Range of lag times}
The  coupled dynamics  (\ref{LJ-SDE}),(\ref{eq:evol_b})  leads to  the
equality of  correlations, as  in (\ref{converged}), for  a particular
value of the lag time $\Delta t=\tau_S,\tau_\perp,\tau_\parallel$.  If
the  parameterized  Markovian  model  is a  good  description  of  the
underlying microscopic behaviour, we should obtain a set of parameters
$\mu_0,a_\alpha,b_\alpha$ that do  not depend on the value  of the lag
time $\Delta t$ used to fit.

To assess  lag dependence and  determine a practical window  of usable
lag  times,  we  repeated  the  parameter  evolution  simulations  for
multiple  choices  of  $\tau_S,\tau_\parallel,$ and  $\tau_\perp$  for
different masses $m_B$ of the heavy particles.  For each lag choice we
ran  $8$  independent  CG   simulations  and  averaged  the  resulting
observables  and  parameter  trajectories to  reduce  estimator  noise
during the evolution stage, using $T_0=T_\alpha=10^{5}$.

From physical considerations, decreasing the tracer mass reduces the
separation of time scales between the tracer and the thermal bath. As
a result, the tracer dynamics becomes less Markovian, which should
manifest as a stronger dependence of the inferred parameters on the
lag time used in the evolution.
\begin{table}[t]
\centering
\begin{tabular}{c c c}
\hline\hline
$m_B$ &  $\tau_S$ &  $\tau_\parallel,\tau_\perp$ \\
\hline
2  & none  & none \\
5  & $[50,2000)$   & $[0.1,0.2]$ \\
10 & $[100,2000)$  & $[0.1,0.5]$ \\
50 & $[500,2000)$  & $[0.1,0.6]$ \\
\hline\hline
\end{tabular}
\caption{Lag-time ranges over which the inferred parameters are
independent of the lag for different tracer masses $m_B$. For
$m_B=2$, no stable regime is observed, reflecting the insufficient
separation of time scales between the tracer and the bath.}
\label{tab:lag_ranges}
\end{table}
The  results in  Table~\ref{tab:lag_ranges} confirm  this expectation.
For the smallest mass $m_B=2$, no stable parameter regime is observed,
consistent with the lack of  time-scale separation between tracers and
bath  particles.   In   this  case,  fitting  a   Markovian  model  to
intrinsically non-Markovian  dynamics leads to parameters  that depend
strongly on  the chosen lag  time $\Delta  t$, indicating that  such a
Markovian  description is  inadequate. For  larger masses,  stable lag
windows emerge beyond the ballistic regime. The lower bound of the MSD
window  increases  approximately  with  tracer  mass,  reflecting  the
correspondingly longer ballistic regime.

Low-Reynolds-number hydrodynamics  predicts that the  mobility depends
only  on  geometric and  solvent  properties,  while the  tracer  mass
affects only  the inertial time  scales of the dynamics.  Although the
mobility extracted  from our  MD simulations does  not follow  the RPY
prediction,  the parameters  obtained  for $m_B=10$  and $m_B=50$  are
nearly   identical,  indicating   that  the   effective  mobility   is
essentially independent of the mass of the heavy particles.

For $m_B=5$, we observe similar but slightly larger coefficients. We
attribute this deviation to a weaker separation of time scales, in
which residual memory and inertial effects are not fully negligible
and are partially absorbed into the effective mobility parameters.

\subsection{Physical picture emerging}

The self-averaging method allows us  to validate the CG Langevin model
for  the heavy  particles and  to extract  several physical  insights,
summarized below:
\begin{itemize}
\item For a mass ratio above approximately $m_B/m_A=5$ the dynamics is
  fairly Markovian, as attested by the fact that the CG Langevin model
  can be  succesfully fitted. A  mass ratio  of $2$ cannot  be fitted,
  which we attribute to  non-Markovian behaviour.
\item Under  the thermodynamic conditions considered  here, long range
  solvent  mediated  interactions  play  a  significant  role  in  the
  dynamics.
\item A pairwise \textit{friction}  representation does not adequately
  describe  the dynamics.  In contrast,  a pairwise  \textit{mobility}
  formulation performs  well, suggesting that the  underlying friction
  is effectively many-body in character.
\item At short separations, solvent layering effects appear to be important,
leading to anisotropic friction.
\item At intermediate separations, the medium induces an approximately
isotropic friction that nevertheless depends on the particle
configuration.
\item At  large separations of the  order of the box  sizes considered
  the friction is significant.
\item  The inferred pair mobility differs qualitatively from the RPY
tensor typically used in colloidal suspension simulations. In
particular, while RPY predicts anisotropic couplings with
$\mu_\parallel/\mu_\perp\to 2$ at large separations, the mobility
inferred from MD is approximately isotropic,
$\mu_\parallel\simeq\mu_\perp$, over intermediate distances
[Fig.~\ref{Fig:LJ-1}(b)]. This difference is not surprising, since
the heavy tracers in the present LJ mixture differ from the bath
particles only through their inertia and do not correspond to rigid
continuum spheres with a prescribed hydrodynamic radius.

\end{itemize}
\section{Conclusion}

We  have  introduced a  self-averaging  framework  for estimating  the
parameters   of  coarse-grained   stochastic  differential   equations
directly from  microscopic trajectory  data.  The  central idea  is to
transform  the parameter  inference problem  into a  dynamical one  by
coupling the  coarse-grained dynamics  to evolution equations  for the
parameters  themselves.   Under  suitable time-scale  separation,  the
resulting extended  dynamics self-averages and converges  to parameter
values  for  which  selected microscopic  and  mesoscopic  observables
coincide. The analysis of  lag-time dependence further illustrates how
the method provides an internal  consistency check for the validity of
the   CG  model   and,   in  particular,   the  underlying   Markovian
approximation.

The method was first validated  in controlled settings. For a Brownian
particle in a  harmonic trap, the procedure  accurately recovered both
static and dynamical parameters and reproduced the full VAFC. A second
validation  using Brownian  particles  with hydrodynamic  interactions
showed  that the  method can  successfully infer  a position-dependent
mobility    tensor   represented    through   a    flexible   B-spline
parametrization.

We then apply the methodology  to a genuine coarse-graining problem: a
Lennard--Jones  binary mixture  in  which heavy  tracer particles  are
immersed  in a  bath of  lighter particles.   From molecular  dynamics
trajectories the  method allows  to infer both  the potential  of mean
force between  the tracers and their  configuration-dependent mobility
tensor.  The resulting coarse-grained model accurately reproduces both
structural  and  dynamical  observables  of  the  microscopic  system,
including the  radial distribution function,  velocity autocorrelation
functions,  and  collective  correlations  across a  range  of  length
scales.   The  resulting Langevin  description  not  only reduces  the
computational cost  relative to  fully atomistic simulations  but also
provides  a physically  transparent representation  of the  system. In
particular, it  offers a  simplified framework that  helps rationalize
and understand the mechanisms governing the dynamics of the underlying
microscopic model.  In  the binary mixture of LJ  particles, the heavy
particles interact  through long  range solvent  mediated interactions
that do not fit into the usual RPY mobility tensor.

Overall, the  proposed self-averaging  approach provides  a systematic
framework for determining both  thermodynamic and transport parameters
of   coarse-grained  stochastic   models  directly   from  microscopic
simulations. By relying only on  averages and correlations of selected
observables,   the   method   avoids  the   explicit   evaluation   of
high-dimensional conditional  expectations and  naturally accommodates
flexible parametrizations  of state-dependent  transport coefficients.
These features  make it a  practical and robust tool  for constructing
physically grounded  coarse-grained models of complex  fluids and soft
matter systems.

\begin{acknowledgments}
This work was supported by grant PID2024-158994OB-C43 funded by MICIU/AEI/10.13039/501100011033. C. M. acknowledges support from predoctoral contract PRE2021-099773, funded by MCIN/AEI/10.13039/501100011033 and by FSE+, associated with project PID2020-117080RB-C54. We acknowledge the computational resources and assistance provided by the Centro de Computación de Alto Rendimiento CCAR-UNED.
\end{acknowledgments}

\appendix

\section{Sign stability criterion}
\label{App:sign}

In this appendix we discuss the selection of the sign $s_Q$ in
(\ref{dalg}) that guarantees the stable convergence of the parameter
dynamics.

Define the time correlation predicted by the mesoscopic model
\begin{align}
  C(\Delta t;\theta)
  &\equiv \llangle Q(a_t)Q(a_{t+\Delta t})\rrangle_\theta,
\end{align}
and the corresponding reference value
\begin{align}
  C_{*}(\Delta t)
  &\equiv \llangle Q(a_0)Q(a_{\Delta t})\rrangle_* .
\end{align}
The matching condition \eqref{C-mes-mic} requires
$C(\Delta t;\theta)=C_{*}(\Delta t)$.
Introducing the mismatch
\begin{equation}
  F_Q(\theta)\equiv C(\Delta t;\theta)-C_{*}(\Delta t),
\end{equation}
the calibration problem amounts to solving the nonlinear equation
$F_Q(\theta)=0$.

Consider one parameter $p$ and the feedback dynamics
\begin{equation}
  \dot p(t)= -\frac{s_Q}{T_{\rm param}}
  \Big[\,C_{*}(\Delta t)-\widehat C(t;\Delta t)\,\Big],
  \qquad s_Q\in\{+1,-1\}.
  \label{eq:dyn-sign-generic}
\end{equation}
Taking the conditional expectation at fixed $p$ yields
\begin{equation}
  \mathbb{E}[\dot p]=
  \frac{s_Q}{T_{\rm param}}\,F_Q(p).
\end{equation}
Linearizing around a solution $p^*$ with $F_Q(p^*)=0$ gives
\begin{equation}
  \delta\dot p=
  \frac{s_Q}{T_{\rm param}}\,F'_Q(p^*)\,\delta p,
  \qquad
  F'_Q(p^*)=
  \frac{\partial C(\Delta t;p)}{\partial p}\bigg|_{p^*}.
\end{equation}
Therefore the fixed point is linearly stable if and only if
\begin{equation}
  s_Q\,F'_Q(p^*)<0,
\end{equation}
which yields the general sign rule
\begin{equation}
 s_Q=-\mathrm{sign}\!\left(
 \frac{\partial C(\Delta t;p)}{\partial p}\bigg|_{p^*}
 \right).
 \label{eq:dyn-sign-rule}
\end{equation}

In words, the sign in the parameter dynamics is determined by the
monotonic response of the chosen time correlation with respect to the
parameter being fitted. If increasing the parameter increases the
correlation, the feedback must act with the opposite sign, and vice
versa.

\subsection{Example 1: Brownian trap}

For the harmonic Langevin oscillator discussed in
Sec.~\ref{B-1D}, increasing the damping coefficient $\gamma$
reduces the momentum correlation at any fixed lag $\tau>0$,
so that
\begin{equation}
  \frac{\partial}{\partial\gamma}C_\gamma(\tau)<0 .
\end{equation}
Applying the stability rule \eqref{eq:dyn-sign-rule} therefore yields
\begin{equation}
  \frac{d\gamma(t)}{dt}
  =-\frac{1}{T_{\rm param}}
  \left[C_p(\tau)-p(t-\tau)p(t)\right].
\end{equation}
If the simulated correlation is too large compared with the target,
$\gamma$ increases and the dynamics becomes more strongly damped,
thereby reducing the correlation.

\subsection{Example 2: Brownian trap with HI}

In the example of Sec.~\ref{N-B-HI} we consider the observables
$X\in\{O,G_\alpha,H_\alpha\}$ with mesoscopic predictions
\begin{equation}
  C_X(\tau;p)\equiv \llangle X(t,\tau)\rrangle_{p}.
\end{equation}
The parameters $\tilde D_0$, $a_\alpha$, and $b_\alpha$ enter the
mobility tensor $\overline{\mathbf D}(\mathbf R)$ and therefore control
the magnitude of the displacement increments
$\Delta\mathbf R_\mu(t,\tau)$.

Increasing any of these mobility coefficients enhances the corresponding
displacement correlations. In particular,
\begin{align}
\partial_{\tilde D_0}\llangle O(t,\tau)\rrangle &>0,\\
\partial_{a_\alpha}\llangle G_\alpha(t,\tau)\rrangle &>0,\\
\partial_{b_\alpha}\llangle H_\alpha(t,\tau)\rrangle &>0 .
\end{align}
Hence $\mathrm{sign}(\partial_p C_X)=+1$, and the stable feedback
sign given by \eqref{eq:dyn-sign-rule} is $s_X=-1$, yielding the update
\begin{equation}
  \dot p(t)=\frac{1}{T_p}\left[ C_{X,*}(\tau)-X(t,\tau)\right].
\end{equation}

\subsection{Example 3: LJ binary mixture}

In the LJ mixture considered in Sec.~\ref{Sec:MD}, the irreversible dynamics is
controlled by the mobility matrix $\boldsymbol{\mu}(\mathbf R)$, while the
friction tensor is its inverse,
$\boldsymbol{\Gamma}(\mathbf R)=\boldsymbol{\mu}(\mathbf R)^{-1}.$
The self parameter $\mu_0$ is fitted through the lagged displacement observable
$O(t,\tau_S)$, whereas the spline coefficients $a_\alpha$ and $b_\alpha$
parameterize the transverse and longitudinal pair mobilities,
\[
\mu_\perp(r)=\sum_\alpha a_\alpha N_{\alpha,p}(r),
\qquad
\mu_\parallel(r)=\sum_\alpha b_\alpha N_{\alpha,p}(r),
\]
and are fitted through the corresponding relative-velocity observables
$G^\perp_\alpha(t,\tau_\perp)$ and
$G^\parallel_\alpha(t,\tau_\parallel)$.

For the self term, increasing $\mu_0$ decreases the effective self-friction and
therefore enhances the tracer displacement at any fixed lag $\tau_S>0$. Hence
\[
\frac{\partial}{\partial\mu_0}\,\langle O(\tau_S)\rangle > 0.
\]
By the general sign rule, this implies $s_O=-1$, which yields the update
\[
\frac{d\mu_0}{dt}
=
\frac{1}{T_0}\Big[\langle O(\tau_S)\rangle_* - O(t,\tau_S)\Big].
\]

For the pair coefficients, the relevant quantity is the mobility of the
\emph{relative} motion of a tracer pair. In a two-particle approximation, the
relative transverse and longitudinal mobilities scale as
\[
\mu_{\rm rel}^\perp(r)\simeq \mu_0-\mu_\perp(r),
\qquad
\mu_{\rm rel}^\parallel(r)\simeq \mu_0-\mu_\parallel(r).
\]
Therefore, increasing $a_\alpha$ increases $\mu_\perp(r)$ over the support of
the basis function $N_{\alpha,p}$ and thus decreases the corresponding relative
transverse mobility; similarly, increasing $b_\alpha$ decreases the relative
longitudinal mobility. Since the observables
$G^\perp_\alpha(t,\tau_\perp)$ and
$G^\parallel_\alpha(t,\tau_\parallel)$ are lagged relative-velocity
correlations, they decrease when the corresponding relative mobility decreases.
Accordingly,
\[
\frac{\partial}{\partial a_\alpha}\,
\big\langle G^\perp_\alpha(\tau_\perp)\big\rangle < 0,
\qquad
\frac{\partial}{\partial b_\alpha}\,
\big\langle G^\parallel_\alpha(\tau_\parallel)\big\rangle < 0.
\]
Applying the sign rule then gives
\[
s_{G^\perp_\alpha}=+1,
\qquad
s_{G^\parallel_\alpha}=+1,
\]
which leads to the stable evolution equations
\begin{align}
\frac{da_\alpha}{dt}
&=
-\frac{1}{T_\alpha}
\Big[\langle G^\perp_\alpha(\tau_\perp)\rangle_*
      -G^\perp_\alpha(t,\tau_\perp)\Big],
\nonumber\\
\frac{db_\alpha}{dt}
&=
-\frac{1}{T_\alpha}
\Big[\langle G^\parallel_\alpha(\tau_\parallel)\rangle_*
      -G^\parallel_\alpha(t,\tau_\parallel)\Big].
\end{align}
Thus, if the simulated relative-velocity correlation is too large compared with
the target, the corresponding coefficient $a_\alpha$ or $b_\alpha$ decreases,
the relative mobility increases less strongly, and the evolution acts to reduce
the discrepancy.

\section{Simulation details}
\label{App:sim-details}
In this appendix we provide the simulation details in the three examples considered in this work.
\subsection{Example 1: Brownian trap}
The stochastic dynamics (\ref{Langevin})  is integrated using the
Grønbech-Jensen and Farago (G-JF)
algorithm       \cite{gronbech-jensen2014},       recommended       in
\cite{Finkelstein2020},  while   the  parameter   evolution  equations
(\ref{dkdg}) are integrated with a first-order Euler scheme due to the
large  value  of  $T_{\rm   param}$.   Unless  otherwise  stated,  the
simulation  parameters  are  $\gamma^\ast  =  k_B  T  =  m  =  1$  and
$dt = 0.001$.  We perform 16 independent runs of  $10^{6}$ steps each,
with $T_{\rm param}=10^{6}$.

\subsection{Example 2: Brownian trap with HI}
For  the ground  truth  (GT)  simulations we  chose  a  system of  100
Brownian   particles   under   the    action   of   the   RPY   tensor
(\ref{RPY-tensor}). This  number of particles  was chosen in  order to
obtain correct  statistics, while  keeping in  mind that  the Cholesky
decomposition of the RPY tensor  becomes increasingly expensive as the
number of particles  grows.  We used the following  parameters for our
simulations:  $D_0=1$,  $a=1$, $d t  =  0.001$,  $l =  5$.   The
observables defined in Eqs.~(\ref{Ottau})--(\ref{Halpha}) are computed
for   each  particle   configuration  and   time-averaged  to   obtain
$\langle \cdot  \rangle_*$. The reported results  are further averaged
over 128 independent runs of the  GT system.  From the GT trajectories
we  found that  the maximum  particle  distance was  around 39  length
units, with a more common maximum distance close to 30 units. We chose
60 basis  for our B-splines,  spanning a range  from 0 to  40 distance
units in order to cover the full range. By construction, the parameter
$b_0$ is set  to zero as it belongs to  a zero inter-particle distance
basis.

For the parameter evolution we also  performed 128 runs, with the same
parameters.  As initial  condition, we  set $\tilde{D}_0=10^{-3}$  and
$a_\alpha=0,b_\alpha=0$ for all $\alpha$. We let the system evolve for
a long period of time until  every parameter in our generic tensor has
converged   to   a    stable   value.   For   the    data   shown   in
Fig.~\ref{fig:rpy_validation_2x2}(c),  we  used  \({\Delta t}  =  0.1\)  and
$\frac{1}{T_0}=10^{-3}$   for  the   parameter  \(\tilde{D}_0\),   and
\({\Delta t}  = 1\)  and $\frac{1}{T_\alpha}=10^{-4}$  for \(a_\alpha\)  and
\(b_\alpha\).
\subsection{Example 3: LJ binary mixture}
All  LAMMPS runs were  carried out at  temperature $T=2$  and number
densities $\rho=0.6$ and $\rho=0.8$ in a periodic cubic box of side length $L=25.6285$,
containing $N_A=10\,000$ light particles ($m_A=1$) and $N_B=100$ heavy
tracers. We considered four tracer masses $m_B \in \{2,5,10,50\}$. All
particles, light and heavy, interact with the same LJ potential with a
shifted  cutoff  $r_c^{\rm  LJ}=3$.   The  equations  of  motion  were
integrated with time step $d t=10^{-3}$ for $2\times10^{6}$ steps
per  trajectory ($2000$  LJ  time  units).  For  each  tracer mass  we
performed $128$ statistically independent runs.  The target quantities
$\langle\cdot\rangle_*$ entering the parameter evolution were obtained
by computing the observables along  each trajectory and averaging over
the ensemble of  runs.  The CG dynamics  (\ref{LJ-SDE}) was integrated
with the G-JF scheme, as in Sec.~\ref{B-1D}.
For  the  conservative  PMF  we   used  $M_U=40$  cubic  B-splines  on
$[r_{\min},r_c^{U}]$ with $r_{\min}=0.83$  and $r_c^{U}=3.0$, matching
the  microscopic LJ  cutoff.  For  the irreversible  mobility we  used
$M_\mu=20$    cubic    B-splines   on    $[r_{\min},r_{\max}]$    with
$r_{\max}=0.49L$. To  mirror  the  microscopic  statistics,  we  perform  $128$
independent   CG  simulations   (as   in  the   MD  ensemble),   using
$d t=10^{-3}$ and trajectories of length $2000$ LJ time units.
\section{B-Splines}
\label{App:Splines}
The  B-spline   basis  functions   are  defined  recursively   over  a
non-decreasing   knot   vector   $\{t_0,t_1,\dots,t_{M+p+1}\}$.    The
zeroth-degree (piecewise constant) basis functions are
\begin{equation}
N_{\alpha,0}(r) =
\begin{cases}
1, & t_\alpha \le r < t_{\alpha+1}, \\
0, & \text{otherwise}.
\end{cases}
\end{equation}
For $p \ge 1$, the basis functions are given by
\begin{align}
N_{\alpha,p}(r) =
&\frac{r - t_\alpha}{t_{\alpha+p} - t_\alpha} \, N_{\alpha,p-1}(r)
  \nonumber\\
  +& \frac{t_{\alpha+p+1} - r}{t_{\alpha+p+1} - t_{\alpha+1}} \, N_{\alpha+1,p-1}(r),
\end{align}
where the fractions are taken to be zero whenever the denominator vanishes.
The basis functions have $p-1$ continuous derivatives (for simple knots), local
support ($N_{\alpha,p}(r)$ is nonzero only on $[t_\alpha,t_{\alpha+p+1})$), and
form a partition of unity: $\sum_{\alpha} N_{\alpha,p}(r)=1$ for all $r$ in the
domain.
\section{Noise-induced drift}
\label{App:NI}

The It\^o drift contains the divergence
\begin{align}
\sum_\nu \frac{\partial}{\partial \mathbf R_\nu}\!\cdot\,
\overline{\mathbf D}_{\mu\nu}.
\end{align}
Since \(\overline{\mathbf D}_{\mu\mu}=\tilde D_0\mathbf I\) is constant, only
the off-diagonal blocks contribute. For \(\mu\neq\nu\),
\begin{align}
\overline{\mathbf D}_{\mu\nu}
=
D_1(R_{\mu\nu})\mathbf I
+
D_{\parallel}(R_{\mu\nu})\mathbf e_{\mu\nu}\mathbf e_{\mu\nu}^T,
\end{align}
with \(\mathbf r_{\mu\nu}=\mathbf R_\mu-\mathbf R_\nu\),
\(R_{\mu\nu}=|\mathbf r_{\mu\nu}|\), and
\(\mathbf e_{\mu\nu}=\mathbf r_{\mu\nu}/R_{\mu\nu}\). Since
\(\partial/\partial \mathbf R_\nu=-\nabla_{\mathbf r_{\mu\nu}}\), one finds
\begin{align}
\frac{\partial}{\partial \mathbf R_\nu}\!\cdot\,
\overline{\mathbf D}_{\mu\nu}
=
-\nabla_{\mathbf r_{\mu\nu}}\!\cdot
\left[
D_1(R_{\mu\nu})\mathbf I
+
D_{\parallel}(R_{\mu\nu})\mathbf e_{\mu\nu}\mathbf e_{\mu\nu}^T
\right].
\end{align}
Using
\begin{align}
\nabla_{\mathbf r}\!\cdot\!\big[f(r)\mathbf I\big] &= f'(r)\mathbf e,
\\
\nabla_{\mathbf r}\!\cdot\!\big[g(r)\mathbf e\mathbf e^T\big]
&=
\left[g'(r)+\frac{2g(r)}{r}\right]\mathbf e
\qquad (d=3),
\end{align}
we obtain
\begin{align}
\frac{\partial}{\partial \mathbf R_\nu}\!\cdot\,
\overline{\mathbf D}_{\mu\nu}
=
-\left[
D_1'(R_{\mu\nu})
+
D_{\parallel}'(R_{\mu\nu})
+
\frac{2D_{\parallel}(R_{\mu\nu})}{R_{\mu\nu}}
\right]\mathbf e_{\mu\nu}.
\end{align}
Therefore,
\begin{align}
&\sum_\nu \frac{\partial}{\partial \mathbf R_\nu}\!\cdot\,
\overline{\mathbf D}_{\mu\nu}
  \nonumber\\
&  =
-\sum_{\nu\neq\mu}
\left[
D_1'(R_{\mu\nu})
+
D_{\parallel}'(R_{\mu\nu})
+
\frac{2D_{\parallel}(R_{\mu\nu})}{R_{\mu\nu}}
\right]\mathbf e_{\mu\nu}.
\end{align}
With
\begin{align}
D_1(r)=\sum_\alpha a_\alpha N_{\alpha,p}(r),
\qquad
D_{\parallel}(r)=\sum_\alpha b_\alpha N_{\alpha,p}(r),
\end{align}
this becomes
\begin{align}
&   \sum_\nu \frac{\partial}{\partial{\bf R}_\nu}\cdot \overline{\mathbf{D}}_{\mu\nu}=
  \nonumber\\
  &
  -\sum_{\nu\neq \mu} \sum_\alpha
  \left[
    (a_\alpha + b_\alpha)\, N'_{\alpha,p}(R_{\mu\nu})
    + \frac{2b_\alpha}{R_{\mu\nu}}\, N_{\alpha,p}(R_{\mu\nu})
  \right]{\bf e}_{\mu\nu}.
\end{align}
which is the expression used in Eq.~(\ref{Noise-induced}).

%\bibliographystyle{apsrev4-2}
%\bibliography{Mibiblioteca}

\begin{thebibliography}{31}%
\makeatletter
\providecommand \@ifxundefined [1]{%
 \@ifx{#1\undefined}
}%
\providecommand \@ifnum [1]{%
 \ifnum #1\expandafter \@firstoftwo
 \else \expandafter \@secondoftwo
 \fi
}%
\providecommand \@ifx [1]{%
 \ifx #1\expandafter \@firstoftwo
 \else \expandafter \@secondoftwo
 \fi
}%
\providecommand \natexlab [1]{#1}%
\providecommand \enquote  [1]{``#1''}%
\providecommand \bibnamefont  [1]{#1}%
\providecommand \bibfnamefont [1]{#1}%
\providecommand \citenamefont [1]{#1}%
\providecommand \href@noop [0]{\@secondoftwo}%
\providecommand \href [0]{\begingroup \@sanitize@url \@href}%
\providecommand \@href[1]{\@@startlink{#1}\@@href}%
\providecommand \@@href[1]{\endgroup#1\@@endlink}%
\providecommand \@sanitize@url [0]{\catcode `\\12\catcode `\$12\catcode
  `\&12\catcode `\#12\catcode `\^12\catcode `\_12\catcode `\%12\relax}%
\providecommand \@@startlink[1]{}%
\providecommand \@@endlink[0]{}%
\providecommand \url  [0]{\begingroup\@sanitize@url \@url }%
\providecommand \@url [1]{\endgroup\@href {#1}{\urlprefix }}%
\providecommand \urlprefix  [0]{URL }%
\providecommand \Eprint [0]{\href }%
\providecommand \doibase [0]{https://doi.org/}%
\providecommand \selectlanguage [0]{\@gobble}%
\providecommand \bibinfo  [0]{\@secondoftwo}%
\providecommand \bibfield  [0]{\@secondoftwo}%
\providecommand \translation [1]{[#1]}%
\providecommand \BibitemOpen [0]{}%
\providecommand \bibitemStop [0]{}%
\providecommand \bibitemNoStop [0]{.\EOS\space}%
\providecommand \EOS [0]{\spacefactor3000\relax}%
\providecommand \BibitemShut  [1]{\csname bibitem#1\endcsname}%
\let\auto@bib@innerbib\@empty
%</preamble>
\bibitem [{\citenamefont {Einstein}(1905)}]{Einstein1905}%
  \BibitemOpen
  \bibfield  {author} {\bibinfo {author} {\bibfnamefont {A.}~\bibnamefont
  {Einstein}},\ }\href@noop {} {\bibfield  {journal} {\bibinfo  {journal} {Ann.
  Phys. (Leipzig)}\ }\textbf {\bibinfo {volume} {19}},\ \bibinfo {pages} {549}
  (\bibinfo {year} {1905})}\BibitemShut {NoStop}%
\bibitem [{\citenamefont {Green}(1952)}]{Green1952}%
  \BibitemOpen
  \bibfield  {author} {\bibinfo {author} {\bibfnamefont {M.}~\bibnamefont
  {Green}},\ }\href@noop {} {\bibfield  {journal} {\bibinfo  {journal} {J.
  Chem. Phys.}\ }\textbf {\bibinfo {volume} {20}},\ \bibinfo {pages} {1281}
  (\bibinfo {year} {1952})}\BibitemShut {NoStop}%
\bibitem [{\citenamefont {Zwanzig}(1961)}]{Zwanzig1961}%
  \BibitemOpen
  \bibfield  {author} {\bibinfo {author} {\bibfnamefont {R.}~\bibnamefont
  {Zwanzig}},\ }\href {https://doi.org/10.1103/PhysRev.124.983} {\bibfield
  {journal} {\bibinfo  {journal} {Physical Review}\ }\textbf {\bibinfo {volume}
  {124}},\ \bibinfo {pages} {983} (\bibinfo {year} {1961})}\BibitemShut
  {NoStop}%
\bibitem [{\citenamefont {Noid}(2023)}]{noid2023}%
  \BibitemOpen
  \bibfield  {author} {\bibinfo {author} {\bibfnamefont {W.~G.}\ \bibnamefont
  {Noid}},\ }\href {https://doi.org/10.1021/acs.jpcb.2c08731} {\bibfield
  {journal} {\bibinfo  {journal} {The Journal of Physical Chemistry B}\
  }\textbf {\bibinfo {volume} {127}},\ \bibinfo {pages} {4174} (\bibinfo {year}
  {2023})}\BibitemShut {NoStop}%
\bibitem [{\citenamefont {Lyubartsev}\ and\ \citenamefont
  {Laaksonen}(1995)}]{lyubartsev1995}%
  \BibitemOpen
  \bibfield  {author} {\bibinfo {author} {\bibfnamefont {A.~P.}\ \bibnamefont
  {Lyubartsev}}\ and\ \bibinfo {author} {\bibfnamefont {A.}~\bibnamefont
  {Laaksonen}},\ }\href {https://doi.org/10.1103/PhysRevE.52.3730} {\bibfield
  {journal} {\bibinfo  {journal} {Physical Review E}\ }\textbf {\bibinfo
  {volume} {52}},\ \bibinfo {pages} {3730} (\bibinfo {year}
  {1995})}\BibitemShut {NoStop}%
\bibitem [{\citenamefont {Izvekov}\ and\ \citenamefont
  {Voth}(2005)}]{izvekov2005}%
  \BibitemOpen
  \bibfield  {author} {\bibinfo {author} {\bibfnamefont {S.}~\bibnamefont
  {Izvekov}}\ and\ \bibinfo {author} {\bibfnamefont {G.~A.}\ \bibnamefont
  {Voth}},\ }\href {https://doi.org/10.1021/jp044629q} {\bibfield  {journal}
  {\bibinfo  {journal} {The Journal of Physical Chemistry B}\ }\textbf
  {\bibinfo {volume} {109}},\ \bibinfo {pages} {2469} (\bibinfo {year}
  {2005})}\BibitemShut {NoStop}%
\bibitem [{\citenamefont {Noid}\ \emph {et~al.}(2008)\citenamefont {Noid},
  \citenamefont {Chu}, \citenamefont {Ayton}, \citenamefont {Krishna},
  \citenamefont {Izvekov}, \citenamefont {a~Voth}, \citenamefont {Das},\ and\
  \citenamefont {Andersen}}]{noid2008}%
  \BibitemOpen
  \bibfield  {author} {\bibinfo {author} {\bibfnamefont {W.~G.}\ \bibnamefont
  {Noid}}, \bibinfo {author} {\bibfnamefont {J.-W.}\ \bibnamefont {Chu}},
  \bibinfo {author} {\bibfnamefont {G.~S.}\ \bibnamefont {Ayton}}, \bibinfo
  {author} {\bibfnamefont {V.}~\bibnamefont {Krishna}}, \bibinfo {author}
  {\bibfnamefont {S.}~\bibnamefont {Izvekov}}, \bibinfo {author} {\bibfnamefont
  {G.}~\bibnamefont {a~Voth}}, \bibinfo {author} {\bibfnamefont
  {A.}~\bibnamefont {Das}},\ and\ \bibinfo {author} {\bibfnamefont {H.~C.}\
  \bibnamefont {Andersen}},\ }\href {https://doi.org/10.1063/1.2938860}
  {\bibfield  {journal} {\bibinfo  {journal} {J. Chem. Phys.}\ }\textbf
  {\bibinfo {volume} {128}},\ \bibinfo {pages} {244114} (\bibinfo {year}
  {2008})}\BibitemShut {NoStop}%
\bibitem [{\citenamefont {Shell}(2008)}]{shell2008}%
  \BibitemOpen
  \bibfield  {author} {\bibinfo {author} {\bibfnamefont {M.~S.}\ \bibnamefont
  {Shell}},\ }\href {https://doi.org/10.1063/1.2992060} {\bibfield  {journal}
  {\bibinfo  {journal} {The Journal of Chemical Physics}\ }\textbf {\bibinfo
  {volume} {129}},\ \bibinfo {pages} {144108} (\bibinfo {year}
  {2008})}\BibitemShut {NoStop}%
\bibitem [{\citenamefont {Papavasiliou}\ \emph {et~al.}(2009)\citenamefont
  {Papavasiliou}, \citenamefont {Pavliotis},\ and\ \citenamefont
  {Stuart}}]{papavasiliou2009a}%
  \BibitemOpen
  \bibfield  {author} {\bibinfo {author} {\bibfnamefont {A.}~\bibnamefont
  {Papavasiliou}}, \bibinfo {author} {\bibfnamefont {G.}~\bibnamefont
  {Pavliotis}},\ and\ \bibinfo {author} {\bibfnamefont {A.}~\bibnamefont
  {Stuart}},\ }\href {https://doi.org/10.1016/j.spa.2009.05.003} {\bibfield
  {journal} {\bibinfo  {journal} {Stochastic Processes and their Applications}\
  }\textbf {\bibinfo {volume} {119}},\ \bibinfo {pages} {3173} (\bibinfo {year}
  {2009})}\BibitemShut {NoStop}%
\bibitem [{\citenamefont {Hij{\'o}n}\ \emph {et~al.}(2010)\citenamefont
  {Hij{\'o}n}, \citenamefont {Espa{\~n}ol}, \citenamefont {{Vanden-Eijnden}},\
  and\ \citenamefont {{Delgado-Buscalioni}}}]{hijón2010}%
  \BibitemOpen
  \bibfield  {author} {\bibinfo {author} {\bibfnamefont {C.}~\bibnamefont
  {Hij{\'o}n}}, \bibinfo {author} {\bibfnamefont {P.}~\bibnamefont
  {Espa{\~n}ol}}, \bibinfo {author} {\bibfnamefont {E.}~\bibnamefont
  {{Vanden-Eijnden}}},\ and\ \bibinfo {author} {\bibfnamefont {R.}~\bibnamefont
  {{Delgado-Buscalioni}}},\ }\href@noop {} {\bibfield  {journal} {\bibinfo
  {journal} {Faraday Discuss.}\ }\textbf {\bibinfo {volume} {144}},\ \bibinfo
  {pages} {301} (\bibinfo {year} {2010})}\BibitemShut {NoStop}%
\bibitem [{\citenamefont {Dequidt}\ and\ \citenamefont
  {Solano~Canchaya}(2015)}]{dequidt2015}%
  \BibitemOpen
  \bibfield  {author} {\bibinfo {author} {\bibfnamefont {A.}~\bibnamefont
  {Dequidt}}\ and\ \bibinfo {author} {\bibfnamefont {J.~G.}\ \bibnamefont
  {Solano~Canchaya}},\ }\href {https://doi.org/10.1063/1.4929557} {\bibfield
  {journal} {\bibinfo  {journal} {The Journal of Chemical Physics}\ }\textbf
  {\bibinfo {volume} {143}},\ \bibinfo {pages} {084122} (\bibinfo {year}
  {2015})}\BibitemShut {NoStop}%
\bibitem [{\citenamefont {Han}\ \emph {et~al.}(2021)\citenamefont {Han},
  \citenamefont {Jin},\ and\ \citenamefont {Voth}}]{han2021}%
  \BibitemOpen
  \bibfield  {author} {\bibinfo {author} {\bibfnamefont {Y.}~\bibnamefont
  {Han}}, \bibinfo {author} {\bibfnamefont {J.}~\bibnamefont {Jin}},\ and\
  \bibinfo {author} {\bibfnamefont {G.~A.}\ \bibnamefont {Voth}},\ }\href
  {https://doi.org/10.1063/5.0035184} {\bibfield  {journal} {\bibinfo
  {journal} {The Journal of Chemical Physics}\ }\textbf {\bibinfo {volume}
  {154}},\ \bibinfo {pages} {084122} (\bibinfo {year} {2021})}\BibitemShut
  {NoStop}%
\bibitem [{\citenamefont {Sokhan}\ and\ \citenamefont
  {Todorov}(2021)}]{sokhan2021}%
  \BibitemOpen
  \bibfield  {author} {\bibinfo {author} {\bibfnamefont {V.~P.}\ \bibnamefont
  {Sokhan}}\ and\ \bibinfo {author} {\bibfnamefont {I.~T.}\ \bibnamefont
  {Todorov}},\ }\href {https://doi.org/10.1080/08927022.2019.1578353}
  {\bibfield  {journal} {\bibinfo  {journal} {Molecular Simulation}\ }\textbf
  {\bibinfo {volume} {47}},\ \bibinfo {pages} {248} (\bibinfo {year}
  {2021})}\BibitemShut {NoStop}%
\bibitem [{\citenamefont {Abdulle}\ \emph {et~al.}(2023)\citenamefont
  {Abdulle}, \citenamefont {Garegnani}, \citenamefont {Pavliotis},
  \citenamefont {Stuart},\ and\ \citenamefont {Zanoni}}]{abdulle2023}%
  \BibitemOpen
  \bibfield  {author} {\bibinfo {author} {\bibfnamefont {A.}~\bibnamefont
  {Abdulle}}, \bibinfo {author} {\bibfnamefont {G.}~\bibnamefont {Garegnani}},
  \bibinfo {author} {\bibfnamefont {G.~A.}\ \bibnamefont {Pavliotis}}, \bibinfo
  {author} {\bibfnamefont {A.~M.}\ \bibnamefont {Stuart}},\ and\ \bibinfo
  {author} {\bibfnamefont {A.}~\bibnamefont {Zanoni}},\ }\href
  {https://doi.org/10.1007/s10208-021-09541-9} {\bibfield  {journal} {\bibinfo
  {journal} {Foundations of Computational Mathematics}\ }\textbf {\bibinfo
  {volume} {23}},\ \bibinfo {pages} {33} (\bibinfo {year} {2023})}\BibitemShut
  {NoStop}%
\bibitem [{\citenamefont {Milster}\ \emph {et~al.}(2025)\citenamefont
  {Milster}, \citenamefont {Dzubiella}, \citenamefont {Stock},\ and\
  \citenamefont {Wolf}}]{milster2025}%
  \BibitemOpen
  \bibfield  {author} {\bibinfo {author} {\bibfnamefont {S.}~\bibnamefont
  {Milster}}, \bibinfo {author} {\bibfnamefont {J.}~\bibnamefont {Dzubiella}},
  \bibinfo {author} {\bibfnamefont {G.}~\bibnamefont {Stock}},\ and\ \bibinfo
  {author} {\bibfnamefont {S.}~\bibnamefont {Wolf}},\ }\href
  {https://doi.org/10.1063/5.0261459} {\bibfield  {journal} {\bibinfo
  {journal} {The Journal of Chemical Physics}\ }\textbf {\bibinfo {volume}
  {162}},\ \bibinfo {pages} {154113} (\bibinfo {year} {2025})}\BibitemShut
  {NoStop}%
\bibitem [{\citenamefont {Dietrich}\ \emph {et~al.}(2023)\citenamefont
  {Dietrich}, \citenamefont {Makeev}, \citenamefont {Kevrekidis}, \citenamefont
  {Evangelou}, \citenamefont {Bertalan}, \citenamefont {Reich},\ and\
  \citenamefont {Kevrekidis}}]{dietrich2023}%
  \BibitemOpen
  \bibfield  {author} {\bibinfo {author} {\bibfnamefont {F.}~\bibnamefont
  {Dietrich}}, \bibinfo {author} {\bibfnamefont {A.}~\bibnamefont {Makeev}},
  \bibinfo {author} {\bibfnamefont {G.}~\bibnamefont {Kevrekidis}}, \bibinfo
  {author} {\bibfnamefont {N.}~\bibnamefont {Evangelou}}, \bibinfo {author}
  {\bibfnamefont {T.}~\bibnamefont {Bertalan}}, \bibinfo {author}
  {\bibfnamefont {S.}~\bibnamefont {Reich}},\ and\ \bibinfo {author}
  {\bibfnamefont {I.~G.}\ \bibnamefont {Kevrekidis}},\ }\href
  {https://doi.org/10.1063/5.0113632} {\bibfield  {journal} {\bibinfo
  {journal} {Chaos: An Interdisciplinary Journal of Nonlinear Science}\
  }\textbf {\bibinfo {volume} {33}},\ \bibinfo {pages} {023121} (\bibinfo
  {year} {2023})}\BibitemShut {NoStop}%
\bibitem [{\citenamefont {Ye}\ \emph {et~al.}(2023)\citenamefont {Ye},
  \citenamefont {Jing},\ and\ \citenamefont {Pan}}]{ye2023}%
  \BibitemOpen
  \bibfield  {author} {\bibinfo {author} {\bibfnamefont {T.}~\bibnamefont
  {Ye}}, \bibinfo {author} {\bibfnamefont {B.}~\bibnamefont {Jing}},\ and\
  \bibinfo {author} {\bibfnamefont {D.}~\bibnamefont {Pan}},\ }\href
  {https://doi.org/10.1016/j.jcp.2022.111857} {\bibfield  {journal} {\bibinfo
  {journal} {Journal of Computational Physics}\ }\textbf {\bibinfo {volume}
  {475}},\ \bibinfo {pages} {111857} (\bibinfo {year} {2023})}\BibitemShut
  {NoStop}%
\bibitem [{\citenamefont {Sachs}\ \emph {et~al.}(2025)\citenamefont {Sachs},
  \citenamefont {Stark}, \citenamefont {Maurer},\ and\ \citenamefont
  {Ortner}}]{sachs2025}%
  \BibitemOpen
  \bibfield  {author} {\bibinfo {author} {\bibfnamefont {M.}~\bibnamefont
  {Sachs}}, \bibinfo {author} {\bibfnamefont {W.~G.}\ \bibnamefont {Stark}},
  \bibinfo {author} {\bibfnamefont {R.~J.}\ \bibnamefont {Maurer}},\ and\
  \bibinfo {author} {\bibfnamefont {C.}~\bibnamefont {Ortner}},\ }\href
  {https://doi.org/10.1088/2632-2153/ada248} {\bibfield  {journal} {\bibinfo
  {journal} {Machine Learning: Science and Technology}\ }\textbf {\bibinfo
  {volume} {6}},\ \bibinfo {pages} {015016} (\bibinfo {year}
  {2025})}\BibitemShut {NoStop}%
\bibitem [{\citenamefont {Kifer}(2001)}]{kifer2001}%
  \BibitemOpen
  \bibfield  {author} {\bibinfo {author} {\bibfnamefont {Y.}~\bibnamefont
  {Kifer}},\ }\href {https://doi.org/10.1142/S0219493701000023} {\bibfield
  {journal} {\bibinfo  {journal} {Stochastics and Dynamics}\ }\textbf {\bibinfo
  {volume} {01}},\ \bibinfo {pages} {1} (\bibinfo {year} {2001})}\BibitemShut
  {NoStop}%
\bibitem [{\citenamefont {Monago}\ \emph {et~al.}(2025)\citenamefont {Monago},
  \citenamefont {Torre}, \citenamefont {{Delgado-Buscalioni}},\ and\
  \citenamefont {Espa{\~n}ol}}]{Monago2025}%
  \BibitemOpen
  \bibfield  {author} {\bibinfo {author} {\bibfnamefont {C.}~\bibnamefont
  {Monago}}, \bibinfo {author} {\bibfnamefont {J.~A. D.~L.}\ \bibnamefont
  {Torre}}, \bibinfo {author} {\bibfnamefont {R.}~\bibnamefont
  {{Delgado-Buscalioni}}},\ and\ \bibinfo {author} {\bibfnamefont
  {P.}~\bibnamefont {Espa{\~n}ol}},\ }\href {https://doi.org/10.1063/5.0255498}
  {\bibfield  {journal} {\bibinfo  {journal} {The Journal of Chemical Physics}\
  }\textbf {\bibinfo {volume} {162}},\ \bibinfo {pages} {114115} (\bibinfo
  {year} {2025})}\BibitemShut {NoStop}%
\bibitem [{\citenamefont {Mason}\ \emph {et~al.}(1997)\citenamefont {Mason},
  \citenamefont {Ganesan}, \citenamefont {{van Zanten}}, \citenamefont
  {Wirtz},\ and\ \citenamefont {Kuo}}]{Mason1997}%
  \BibitemOpen
  \bibfield  {author} {\bibinfo {author} {\bibfnamefont {T.~G.}\ \bibnamefont
  {Mason}}, \bibinfo {author} {\bibfnamefont {K.}~\bibnamefont {Ganesan}},
  \bibinfo {author} {\bibfnamefont {J.~H.}\ \bibnamefont {{van Zanten}}},
  \bibinfo {author} {\bibfnamefont {D.}~\bibnamefont {Wirtz}},\ and\ \bibinfo
  {author} {\bibfnamefont {S.~C.}\ \bibnamefont {Kuo}},\ }\href
  {https://doi.org/10.1103/PhysRevLett.79.3282} {\bibfield  {journal} {\bibinfo
   {journal} {Phys. Rev. Lett.}\ }\textbf {\bibinfo {volume} {79}},\ \bibinfo
  {pages} {3282} (\bibinfo {year} {1997})}\BibitemShut {NoStop}%
\bibitem [{\citenamefont {Waigh}(2005)}]{waigh2005}%
  \BibitemOpen
  \bibfield  {author} {\bibinfo {author} {\bibfnamefont {T.~A.}\ \bibnamefont
  {Waigh}},\ }\href {https://doi.org/10.1088/0034-4885/68/3/R04} {\bibfield
  {journal} {\bibinfo  {journal} {Reports on Progress in Physics}\ }\textbf
  {\bibinfo {volume} {68}},\ \bibinfo {pages} {685} (\bibinfo {year}
  {2005})}\BibitemShut {NoStop}%
\bibitem [{\citenamefont {Bocquet}\ \emph {et~al.}(1997)\citenamefont
  {Bocquet}, \citenamefont {Hansen},\ and\ \citenamefont
  {Piasecki}}]{bocquet1997}%
  \BibitemOpen
  \bibfield  {author} {\bibinfo {author} {\bibfnamefont {L.}~\bibnamefont
  {Bocquet}}, \bibinfo {author} {\bibfnamefont {J.-P.}\ \bibnamefont
  {Hansen}},\ and\ \bibinfo {author} {\bibfnamefont {J.}~\bibnamefont
  {Piasecki}},\ }\href {https://doi.org/10.1007/BF02770768} {\bibfield
  {journal} {\bibinfo  {journal} {Journal of Statistical Physics}\ }\textbf
  {\bibinfo {volume} {89}},\ \bibinfo {pages} {321} (\bibinfo {year}
  {1997})}\BibitemShut {NoStop}%
\bibitem [{\citenamefont {Reith}\ \emph {et~al.}(2001)\citenamefont {Reith},
  \citenamefont {Meyer},\ and\ \citenamefont
  {{M{\"u}ller-Plathe}}}]{Reith2001}%
  \BibitemOpen
  \bibfield  {author} {\bibinfo {author} {\bibfnamefont {D.}~\bibnamefont
  {Reith}}, \bibinfo {author} {\bibfnamefont {H.}~\bibnamefont {Meyer}},\ and\
  \bibinfo {author} {\bibfnamefont {F.}~\bibnamefont {{M{\"u}ller-Plathe}}},\
  }\href {https://doi.org/10.1021/ma001499k} {\bibfield  {journal} {\bibinfo
  {journal} {Macromolecules}\ }\textbf {\bibinfo {volume} {34}},\ \bibinfo
  {pages} {2335} (\bibinfo {year} {2001})},\ \Eprint
  {https://arxiv.org/abs/cond-mat/0008338} {arXiv:cond-mat/0008338}
  \BibitemShut {NoStop}%
\bibitem [{\citenamefont {Press}\ \emph {et~al.}(1992)\citenamefont {Press},
  \citenamefont {Teukolski}, \citenamefont {Vetterling},\ and\ \citenamefont
  {Flannery}}]{Press1992}%
  \BibitemOpen
  \bibfield  {author} {\bibinfo {author} {\bibfnamefont {W.}~\bibnamefont
  {Press}}, \bibinfo {author} {\bibfnamefont {S.}~\bibnamefont {Teukolski}},
  \bibinfo {author} {\bibfnamefont {W.}~\bibnamefont {Vetterling}},\ and\
  \bibinfo {author} {\bibfnamefont {B.}~\bibnamefont {Flannery}},\ }\href@noop
  {} {\emph {\bibinfo {title} {Numerical Recipes in Fortran 77}}}\ (\bibinfo
  {publisher} {Cambridge University Press},\ \bibinfo {year}
  {1992})\BibitemShut {NoStop}%
\bibitem [{\citenamefont {Givon}\ \emph {et~al.}(2004)\citenamefont {Givon},
  \citenamefont {Kupferman},\ and\ \citenamefont {Stuart}}]{Givon2004}%
  \BibitemOpen
  \bibfield  {author} {\bibinfo {author} {\bibfnamefont {D.}~\bibnamefont
  {Givon}}, \bibinfo {author} {\bibfnamefont {R.}~\bibnamefont {Kupferman}},\
  and\ \bibinfo {author} {\bibfnamefont {A.}~\bibnamefont {Stuart}},\ }\href
  {https://doi.org/10.1088/0951-7715/17/6/R01} {\bibfield  {journal} {\bibinfo
  {journal} {Nonlinearity}\ }\textbf {\bibinfo {volume} {17}},\ \bibinfo
  {pages} {R55} (\bibinfo {year} {2004})}\BibitemShut {NoStop}%
\bibitem [{\citenamefont {Beard}\ and\ \citenamefont
  {Schlick}(2000)}]{beard2000}%
  \BibitemOpen
  \bibfield  {author} {\bibinfo {author} {\bibfnamefont {D.~A.}\ \bibnamefont
  {Beard}}\ and\ \bibinfo {author} {\bibfnamefont {T.}~\bibnamefont
  {Schlick}},\ }\href {https://doi.org/10.1063/1.481331} {\bibfield  {journal}
  {\bibinfo  {journal} {The Journal of Chemical Physics}\ }\textbf {\bibinfo
  {volume} {112}},\ \bibinfo {pages} {7313} (\bibinfo {year}
  {2000})}\BibitemShut {NoStop}%
\bibitem [{\citenamefont {Tanygin}\ and\ \citenamefont
  {Melchionna}(2024)}]{tanygin2024}%
  \BibitemOpen
  \bibfield  {author} {\bibinfo {author} {\bibfnamefont {B.}~\bibnamefont
  {Tanygin}}\ and\ \bibinfo {author} {\bibfnamefont {S.}~\bibnamefont
  {Melchionna}},\ }\href {https://doi.org/10.1016/j.cpc.2024.109152} {\bibfield
   {journal} {\bibinfo  {journal} {Computer Physics Communications}\ }\textbf
  {\bibinfo {volume} {299}},\ \bibinfo {pages} {109152} (\bibinfo {year}
  {2024})}\BibitemShut {NoStop}%
\bibitem [{\citenamefont {Thompson}\ \emph {et~al.}(2022)\citenamefont
  {Thompson}, \citenamefont {Aktulga}, \citenamefont {Berger}, \citenamefont
  {Bolintineanu}, \citenamefont {Brown}, \citenamefont {Crozier}, \citenamefont
  {In~'T~Veld}, \citenamefont {Kohlmeyer}, \citenamefont {Moore}, \citenamefont
  {Nguyen}, \citenamefont {Shan}, \citenamefont {Stevens}, \citenamefont
  {Tranchida}, \citenamefont {Trott},\ and\ \citenamefont
  {Plimpton}}]{thompson2022}%
  \BibitemOpen
  \bibfield  {author} {\bibinfo {author} {\bibfnamefont {A.~P.}\ \bibnamefont
  {Thompson}}, \bibinfo {author} {\bibfnamefont {H.~M.}\ \bibnamefont
  {Aktulga}}, \bibinfo {author} {\bibfnamefont {R.}~\bibnamefont {Berger}},
  \bibinfo {author} {\bibfnamefont {D.~S.}\ \bibnamefont {Bolintineanu}},
  \bibinfo {author} {\bibfnamefont {W.~M.}\ \bibnamefont {Brown}}, \bibinfo
  {author} {\bibfnamefont {P.~S.}\ \bibnamefont {Crozier}}, \bibinfo {author}
  {\bibfnamefont {P.~J.}\ \bibnamefont {In~'T~Veld}}, \bibinfo {author}
  {\bibfnamefont {A.}~\bibnamefont {Kohlmeyer}}, \bibinfo {author}
  {\bibfnamefont {S.~G.}\ \bibnamefont {Moore}}, \bibinfo {author}
  {\bibfnamefont {T.~D.}\ \bibnamefont {Nguyen}}, \bibinfo {author}
  {\bibfnamefont {R.}~\bibnamefont {Shan}}, \bibinfo {author} {\bibfnamefont
  {M.~J.}\ \bibnamefont {Stevens}}, \bibinfo {author} {\bibfnamefont
  {J.}~\bibnamefont {Tranchida}}, \bibinfo {author} {\bibfnamefont
  {C.}~\bibnamefont {Trott}},\ and\ \bibinfo {author} {\bibfnamefont {S.~J.}\
  \bibnamefont {Plimpton}},\ }\href {https://doi.org/10.1016/j.cpc.2021.108171}
  {\bibfield  {journal} {\bibinfo  {journal} {Computer Physics Communications}\
  }\textbf {\bibinfo {volume} {271}},\ \bibinfo {pages} {108171} (\bibinfo
  {year} {2022})}\BibitemShut {NoStop}%
\bibitem [{\citenamefont {{Gr{\o}nbech-Jensen}}\ \emph
  {et~al.}(2014)\citenamefont {{Gr{\o}nbech-Jensen}}, \citenamefont {Hayre},\
  and\ \citenamefont {Farago}}]{gronbech-jensen2014}%
  \BibitemOpen
  \bibfield  {author} {\bibinfo {author} {\bibfnamefont {N.}~\bibnamefont
  {{Gr{\o}nbech-Jensen}}}, \bibinfo {author} {\bibfnamefont {N.~R.}\
  \bibnamefont {Hayre}},\ and\ \bibinfo {author} {\bibfnamefont
  {O.}~\bibnamefont {Farago}},\ }\href
  {https://doi.org/10.1016/j.cpc.2013.10.006} {\bibfield  {journal} {\bibinfo
  {journal} {Computer Physics Communications}\ }\textbf {\bibinfo {volume}
  {185}},\ \bibinfo {pages} {524} (\bibinfo {year} {2014})}\BibitemShut
  {NoStop}%
\bibitem [{\citenamefont {Finkelstein}\ \emph {et~al.}(2020)\citenamefont
  {Finkelstein}, \citenamefont {Fiorin},\ and\ \citenamefont
  {Seibold}}]{Finkelstein2020}%
  \BibitemOpen
  \bibfield  {author} {\bibinfo {author} {\bibfnamefont {J.}~\bibnamefont
  {Finkelstein}}, \bibinfo {author} {\bibfnamefont {G.}~\bibnamefont
  {Fiorin}},\ and\ \bibinfo {author} {\bibfnamefont {B.}~\bibnamefont
  {Seibold}},\ }\href {https://doi.org/10.1080/00268976.2019.1649493}
  {\bibfield  {journal} {\bibinfo  {journal} {Molecular Physics}\ }\textbf
  {\bibinfo {volume} {118}},\ \bibinfo {pages} {e1649493} (\bibinfo {year}
  {2020})}\BibitemShut {NoStop}%
\end{thebibliography}

\end{document}